\documentclass[aps,prd,superscriptaddress,showpacs,twocolumn,floatfix]{revtex4}
\usepackage{graphicx}

\newcommand{\bi}[1]{\bibitem{#1}}

\newcommand{\ba}{\begin{eqnarray}}
\newcommand{\ea}{\end{eqnarray}}
\newcommand{\beqs}{\begin{eqnarray}}
\newcommand{\eeqs}{\end{eqnarray}}

\begin{document}
\title{Nucleon structure and  high energy interactions}

\author{  O.V. Selyugin\footnote{selugin@theor.jinr.ru} } 
\address{\it Bogoliubov Laboratory of Theoretical Physics, \\
Joint Institute for Nuclear Research,
141980 Dubna, Moscow region, Russia }

\begin{abstract}
  On the basis of the representation of the generalized structure of nucleons
   a new model of the hadron interaction at high energies is presented.
The new  t-dependence of the generalized parton distributions (GPDs)
is obtained from
the comparative analysis of different sets of the  parton distribution functions (PDFs),
    based on the description of the whole sets of experimental data
    of electromagnetic form factors of the proton and neutron. 
   Taking into account
  the  different moments of 
  GPDs of the hadron  the quantitative descriptions   of all
  existing experimental data of the proton-proton and proton-antiproton elastic
    scattering from  $\sqrt{s} = 9.8$ GeV to 
  $  8 \ $ TeV, including
  the Coulomb range and large momentum transfers up to $-t=15$ GeV$^2$, are obtained with a few
   free fitting high energy parameters.
  The real part of the hadronic elastic scattering amplitude is determined only
   through complex $s$ satisfying the dispersion relations.
The negligible contributions of the
  hard Pomeron and the presence of the non-small contributions of the maximal Odderon are obtained.
   The non-dying   form of the spin-flip amplitude is examined as well.
   The structure of the Born term and unitarized scattering amplitude are analysed.
   It is shown that the Black Disk Limit for the elastic scatering amplitude
   is not reached at LHC energies.
  Predictions for LHC energies are made.
\end{abstract}
%
%
\maketitle %

\section{\label{sec:intro}Introduction}

    A topical problem of  modern physics of elementary particles,
    the exploring of the dynamics of the strong interaction processes
    at high energies, is considered in the framework of different approaches
    using various models of the structure of hadrons and the dynamics of their interactions.
    Different models for the description of hadron interaction at large distances are developed.
    They are based on the general quantum field theory principles (analyticity, unitarity, and so on).
   The relativistic models of high energy scattering based on the quasipotential
    approach \cite{Matveev,Brodsky}
   occupy an important place among them. Here, the hypothesis
   about the existence of the local smooth quasipotential
   giving an adequate description of  high energy scattering processes   is essential.
   In the region of small angles of the scattering the eikonal approach can be used
    as a consequence of the  smoothness of the quasipotential \cite{Brodsky-ei}.
    The smoothness of the quasipotential is related to the dynamics
   of  two-particle interactions and means that at high energies the hadrons behave
   as loose extended objects with finite dimensions.


    The elastic hadron-hadron scattering plays an important role in the investigation
    of the strong interaction. For the description of the interaction at small distances
    there is the exact theory, QCD, but for the interaction at large distances,
    which is the basis for the elastic scattering at small angles, the calculation
    in the framework of QCD is impossible at present. These two domains are tightly connected
    with the experimental determination of the parameters of the elastic scattering
    and are very important for the development of the modern strong interaction theory \cite{Jacob}.

   Only in the region of small angles
  the basic properties of
  the non-perturbative strong interaction: the total cross section,
  the slope of the diffraction peak and the parameter $\rho(s,t) $ - ratio of the real part to the imaginary
  part of the scattering amplitude,  can be measured.
  Their values
  are connected, on the one hand, with the large-scale structure of hadrons and,
  on the other hand, with the first principles which lead to the
  theorems on the behavior of the scattering amplitudes at asymptotic
  energies \cite{mart,roy}.

There are indeed many different models for the description of hadron elastic
 scattering at small angles \cite{Rev-LHC,TOTEM-1395}.  They lead to  the different
 predictions for the structure of the scattering amplitude at asymptotic
 energies, where the diffraction  processes can display complicated
 features \cite{dif04}.  This concerns especially the asymptotic unitarity
 bound connected with the Black Disk Limit (BDL) \cite{CPS-EPJ08} and the influence of the saturation regime on the
 differential cross sections \cite{CS-PRL09}.

   In the Chow-Yang model \cite{WY-65,CY-68} it was assumed
   that the hadron interaction  is    proportional
  to the overlapping of the matter distribution of the hadrons,
   and  Wu and Yang  \cite{WY-65} suggested
   that the matter distribution  is proportional to the charge distribution of the hadron.
 Many models  used  the electromagnetic form factors of the hadron
but, in most part, they changed its form to describe the experimental data,
 as was made in  the famous  Bourrely-Soffer-Wu model \cite{BSW}.
 The parameters of the obtained form-factor are determined by the fit of the differential cross sections.
 The authors noted that the form factor is "parameterized like an electromagnetic form factor, as  two poles,
 and the slowly varying function reflects
 the approximate proportionality between the charge density
 and hadronic matter distribution inside a proton."

   In  paper \cite{Miettinen}, it was proposed that the  hadron form factor is proportional to the matter
   distribution. The matter distributions in the hadron are tightly connected with the energy momentum tensor
   \cite{Pagels}.  In \cite{Broniow}, it was noted that "the gravitational form factors are related to the matrix
   elements of the energy-momentum tensor in a hadronic state,
   thus providing the distribution of matter within the hadron".
   The recent picture of the hadron structure is determined
   by the general parton distributions (GPDs) \cite{Ji97,R97}
   which  include as part the parton distribution functions (PDFs). 
   The first and second moments of GPDs give  two hadron form factors.

     Before the introduction of GPDs
      a similar representation for the scattering amplitude was used
      in work of S. Sanielevici and P. Valin (1984) \cite{Valin}
\ba
  K_{p}(q^2) = \frac{1}{3}  \int^{1}_{0} \ dx \ x [2L^{U}_{p}(x) \ T^{U}_{p}(\vec{k}) +
  2L^{D}_{p}(x) \ T^{D}_{p}(\vec{k});  \nonumber
\ea
   Here the function $L^{U,D}_{p}(x) $ represents the parton distributions and $T^{U,D}_{p}(\vec{k})$
   represents the transfer momentum dependence.
  Note that the whole structure of the amplitude  corresponds  to the second moment of GPDs.

 Usually, models of high energy hadron interaction include a different kind of
 leading Reggions: one or a few pomerons (including the soft and hard pomerons), the odderons
  with intercept equal to the pomeron (maximal odderon) or with intercept close (or less) unity
  and sometimes  the spin-flip amplitude.
  The effect of the hard pomeron contribution on the elastic differential cross sections is very important for  understanding the properties of QCD in the non-perturbative regime \cite{mrt}.
  Note that $\rho(s,t)$ of the hard pomeron
  is essentially larger than  $\rho(s,t)$  of the soft pomeron.
  In \cite{DL-11hp}, it is suggested that such a contribution can explain the preliminary result
   of the TOTEM Collaboration \cite{TOTEM-111008a} on the elastic proton-proton differential cross sections.

   In our high energy general structure (HEGS) model \cite{HEGS-JEP12},
   the real part of the hadronic amplitude is determined only through complex $s$ satisfying the cross symmetric relation.  In the framework of the model, the quantitative  description of all  existing experimental data
   at $52.8 \leq \sqrt{s} \leq 1960  $ GeV,
   including   the Coulomb range and large momentum transfers $0.0008 \leq |t| \leq  9.75 \ $GeV$^2$ ,
    is obtained with only   $3$ fitting high energy parameters.
    The comparison of the predictions of the model  at  $7$ TeV and
    preliminary data of the TOTEM collaboration are shown to coincide well.
        In  \cite{NP-HP}, the contribution  of the hard pomeron in the elastic scattering at small angles at
        high energies was examined. It was found that such a contribution is invisible
        in the existing experimental data  including the new data of LHC.
        In  \cite{DL-13hp}, they came to the same result.

 In the framework of the model, only the Born term of the scattering amplitude
   is introduced. Then the whole scattering amplitude is obtained
    as a result of the unitarization procedure  of the hadron Born term that is then  summed with the Coulomb term.
     The Coulomb-hadron interference phase is also taken into account.
      The essential moment of the model is that both parts of the Born term
  of the scattering amplitude have the positive sign,  and the diffraction structure is determined by the unitarization procedure.

  Now we present the extended variant of the HEGS model \cite{HEGS-JEP12},
 based on the assumption that the hadron interaction is sensitive
 to the generalized parton distributions (GPDs),
  whose moments can be represented 
  in the form of  two different distributions:
   charge and matter, separately.
  Hence, this model used the exact electromagnetic and
   matter form-factors determined by one   function - generalized parton distributions (GPDs).
     Both the form factors are independent of the fitting procedure
    of the differential cross sections  of the elastic hadron scattering.
     Note that the form of  GPDs is determined,
      on  the one hand, by   the deep-inelastic processes and,
      on the other hand, by the measure of the electromagnetic form factor
      from  the electron-nucleon elastic scattering.
 We support this picture   by a good description of the experimental data
 in the Coulomb-hadron interference region and large momentum transfer
  at high energies by one amplitude with a few free parameters.

              The differential cross
  sections of nucleon-nucleon elastic scattering  can be written as the sum of different
  helicity  amplitudes:
\begin{eqnarray}
  \frac{d\sigma}{dt} =
 \frac{2 \pi}{s^{2}} (|\Phi_{1}|^{2} +|\Phi_{2}|^{2} +|\Phi_{3}|^{2}
  +|\Phi_{4}|^{2}
  +4 | \Phi_{5}|^{2} ). \label{dsdt}
\end{eqnarray}
\renewcommand{\bottomfraction}{0.7}
  The total helicity amplitudes can be written as $\Phi_{i}(s,t) =
  F^{h}_{i}(s,t)+F^{\rm em}_{i}(s,t) e^{\varphi(s,t)} $\,, where
 $F^{h}_{i}(s,t) $ comes from the strong interactions,
 $F^{\rm em}_{i}(s,t) $ from the electromagnetic interactions and
 $\varphi(s,t) $
 is the interference phase factor between the electromagnetic and strong
 interactions \cite{selmp1,selmp2,Selphase}.

 The structure of the paper is as follows:
first, in Section 2 
  the basis of the first variant of the  
  HEGS  model   is discussed shortly.

     In Section 3, the hadron form factors are analyzed by the new form of
     the general parton distributions (GPDs) with taking into account two forms of PDFs,
     which give the best descriptions of the electromagnetic form factors of the proton and neutron.
     As a result, the new forms of the electromagnetic and matter form factors are obtained.

 Section 4 is devoted to the study   of all  existing experimental data
  of the proton-proton and proton-antiproton elastic scattering
  from  $\sqrt{s} = 9.8$ GeV to 
  $  8 \ $ TeV, including
  the Coulomb range and large momentum transfers up to $-t=15$ GeV$^2$
  in the framework of the model  with a few
   free fitting high energy parameters.

  In section 5,   the structure of the
  elastic hadron scattering amplitude, obtained in the framework of the model, is analyzed.
  First, the Born term of the scattering amplitude
  is discussed with its  cross-even and cross-odd separate parts.
  Then  the obtained  overlapping function
  is considered  in the impact parameter representation.
  And, finally, the full form of the scattering amplitude, obtained after
  integration over the impact parameter, is considered.
  Especially, we examine the energy and momentum transfer dependence
  of the slope of the scattering amplitude as the Born term and as the full term of the scattering
  amplitude.
  Finally, the  obtained results and the comparison with other models and some predictions
  of  our model are discussed in  Section 6.

\section{The High Energy General Structure (HEGS)  Model }
The model is based on the representation that at high energies the hadron interaction in the non-perturbative regime
      is determined by the reggenized-gluon exchange. The cross-even part of this amplitude can have two  non-perturbative parts, possible standard pomeron - $P_{2g}$ and the  cross-even part of  three non-perturbative gluons $P_{3g}$.
      The interaction of these two objects is proportional to two different form factors of the hadron.
      This is the main assumption of the model. Of course, we cannot insist on the origin of the second
      term of the scattering amplitude.
       However, in any case, it has the cross-even properties, positive sign and the slope is the same as for the odderon.
        The second important assumption is that we chose the slope of the second term  four
        times smaller than the slope of the first term.  
      All  terms have the same intercept.

      The form factors are determined by the general parton distributions of the hadron (GPDs) \cite{StrFF-PN14}.
      The first form factor, corresponding to the first momentum of GPDs is the standard electromagnetic
      form factor - $G(t)$. The second form factor is determined by the second momentum of GPDs -$A(t)$.
      The parameters and $t$-dependence of the GPDs are determined by the standard parton distribution
      functions, so by the experimental data on the deep inelastic scattering and by the experimental data
      for the electromagnetic form factors (see \cite{GPD-ST-PRD09}).

           The electromagnetic form factors can be represented as the first  moments of GPDs
           with $\xi=0 $
\ba
 F_{1}(t) &=& \int^{1}_{0}  dx  \ \sum_{u,d} {\cal{ H}}^{q} (x, t);  \\ \nonumber
 F_{2} (t) &=& \int^{1}_{0} dx \ \sum_{u,d} {\cal{E}}^{q} (x,  t),
\ea
  In \cite{GPD-ST-PRD09}
 the $t$-dependence of  GPDs  in the form
\ba
{\cal{H}}^{q} (x,t) \  &=& q(x)_{nf} \   exp [  a_{+}  \
\frac{(1-x)^2}{x^{m} } \ t ];  \\  \nonumber
{\cal{E}}^{q} (x,t) \  &=& q(x)_{sf} \   exp [  a_{-}  \
\frac{(1-x)^2}{x^{m} } \ t ];
\label{GPD0}
\ea
 was researched.
The function $q(x)$ was chosen at the same scale $\mu^2=1$ as in \cite{R04},
which is based on the MRST2002 global fit \cite{MRST02}.

 For $\xi=0 $ one has the second moment of GPDs
\ba
\int^{1}_{0} \ dx \ x \sum_{u,d}[{\cal{H}}(x,t) \pm {\cal{E}}(x,t)] = A_{h}(t) \pm B_{h}(t).
\ea
The integration of the second moment of GPDs over $x$ give  the momentum-transfer representation
  of the form factor. 
  It was  approximated  by the dipole form \cite{HEGS-JEP12}
 $ A(t)=L^{4}_{2}/(L^{2}_{2}-t)^2$.   

      Hence, the Born term of the elastic hadron amplitude can be written as
  \begin{eqnarray}
 F_{h}^{Born}(s,t) \ =&&  h_1 \ G^{2}(t) \ F_{a}(s,t) \ (1+r_1/\hat{s}^{0.5})  \\ \nonumber
    \   && +  h_{2} \  A^{2}(t) \ F_{b}(s,t) \ (1+r_2/\hat{s}^{0.5}),
    \label{FB}
\end{eqnarray}
  where $F_{a}(s,t)$ and $F_{b}(s,t)$  have the standard Regge form 
  \begin{eqnarray}
 F_{a}(s,t) \ = \hat{s}^{\epsilon_1} \ e^{B(s) \ t}, \ \ \
 F_{b}(s,t) \ = \hat{s}^{\epsilon_1} \ e^{B(s)/4 \ t}.
\label{FB-ab}
\end{eqnarray}
  The slope of the scattering amplitude has the  logarithmic dependence on the energy,
 $   B(s) = \alpha^{\prime} \ ln(\hat{s})$,
  with $\alpha^{\prime}=0.24$ GeV$^{-2}$ and $\hat{s}=s e^{-i \pi/2}/s_{0},  \ \ \ s_{0}=1 \ {\rm GeV^2} $.
 The final elastic  hadron scattering amplitude is obtained after unitarization of the  Born term.
    So, first, we have to calculate the eikonal phase
   \begin{eqnarray}
 \chi(s,b) \   = -\frac{1}{2 \pi}
   \ \int \ d^2 q \ e^{i \vec{b} \cdot \vec{q} } \  F^{\rm Born}_{h}\left(s,q^2\right)\,
 \label{chi}
 \end{eqnarray}
  and then obtain the final hadron scattering amplitude
    \begin{eqnarray}
 F_{h}(s,t) = i s
    \ \int \ b \ J_{0}(b q)  \ \Gamma(s,b)   \ d b\,  \ \ \  \ \ \
 \label{FAmpl}
 \end{eqnarray}
 with
     \begin{eqnarray}
  \Gamma(s,b)  = 1- \exp[\chi(s,b)].
 \label{Gamma}
\end{eqnarray}

%

\begin{table}
 \caption{Experimental data of the electromagnetic form factor)}
\label{Table-2}
\begin{center}
\begin{tabular}{|l|c|c|} \hline    \hline
 $N$ points \ \ \ \ \ \ \ \ \  & \ \ \ \  Proton  \ \ \ \  &   \ \ \ \ References  \ \ \ \          \\
 111 & $G^{p}_{E} $ &  [37-43]  \\
 196  & $G^{p}_{M}$  &  [37,39,40,44-46] \\
  & & [38-43]
    \\
87 & $\mu G^{p}_{E}/G^{p}_{M}  $  &  [38,39,44,47-49]  \\
    &              &             \\   \hline

   &   neutron      &                \\
13 & $G^{n}_{E} $ &  [50-56]  \\
         & &    [57,58]
                \\
38 & $G^{n}_{M}$  & [59-63]
  \\
6 & $\mu G^{n}_{E}/G^{n}_{M}  $  &  [52,64]  \\
   &              &             \\
 \hline     \hline
\end{tabular}
\end{center}
  \end{table}

 The model has only three high-energy fitting parameters
 and two low-energy parameters, which reflect some small contribution
 coming from the different low-energy terms.

  We take all existing
      experimental data in the energy range $52.8 \leq \sqrt{s} \leq 1960 \ $GeV
      and the region of the momentum transfer $0.0008 \leq \ -t \ \leq 9.75 \ $GeV$^2$
      of the elastic differential cross sections of proton-proton and proton-antiproton
      data \cite{data-Sp}.     
       So we include the whole Coulomb-hadron interference region where the experimental errors are       remarkably small.

As a result, one obtains $\sum \chi^2_i /N \simeq 1.8$, where $N=975$ is
the number of experimental points.
Note that the parameters of the model are energy independent.
The energy dependence of the scattering amplitude is determined
only by the single intercept and the logarithmic dependence on $s$ of the slope.

    In the framework of this model the quantitative  description of all
  existing experimental data at $52.8  \leq \sqrt{s} \leq  1960 \ $GeV, including
  the Coulomb range and large momentum transfers ($0.0008 \leq |t| \leq  9.75 \ $GeV$^2$), is obtained
   with only  three   high-energy fitting  parameters. Hence, the model is very sensitive to any additional
   contribution.

%
\begin{table*}
 \caption{The fitting parameters of the GPDs with flavor dependence}
\label{Table-3}
\begin{center}
\begin{tabular}{|c|c|c|c|c|c|c|c|c|c||c|} \hline
          &               &               &       &   &   & & & & &   \\
 Model  & $\epsilon_{u}$& $\epsilon_{d}$ &$\epsilon_{0}$ & $m$ &  $\alpha_{H}$ & $\alpha_{E}$  & $x_0$ &
  $z_{u}$ & $z_{d}$ &
    $ \chi^{2}_{+4p}/\chi^2_{0}$   \\
  & $\pm0.02$ &  $\pm0.01$ &  $\pm0.01$&  $\pm0.01$ & $\pm0.03$   &$\pm0.07$ & $\pm0.002$& $\pm0.03$& $\pm0.03$ & \\ \hline
         &      &     &         &     &        &         &          &       &        & \\
  ABKM09 & 0.11 & 0.2 & $0.09 $ &0.42 &$0.45$  & $0.57$  & 0.004    &$0.67$ &-1.88   &$0.91$  \\
  ABM12  & 0.13 & 0.04& $0.13 $ &0.41 &$0.47$  & $0.60$  & $0.002$  &$0.54$ &$-2.06$ &$0.8

  9$ \\
         &      &     &         &     &        &         &          &       &        &        \\
 \hline
\end{tabular}
\end{center}
  \end{table*}

\begin{figure}
\includegraphics[width=.4\textwidth]{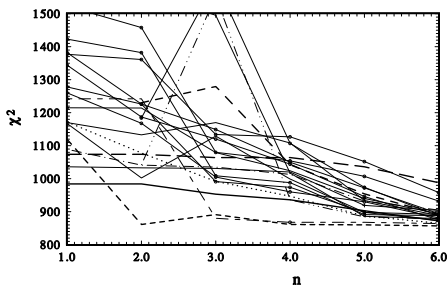} 
\caption{The sum of $\chi^{2}_{i}$ of the descriptions of the proton and neutron electromagnetic form factors by different PDFs over an increasing number of free parameters [Eqs. (10) and  (11)]. 
  }
\label{Fig1}
\end{figure}

\begin{figure}
\includegraphics[width=.4\textwidth]{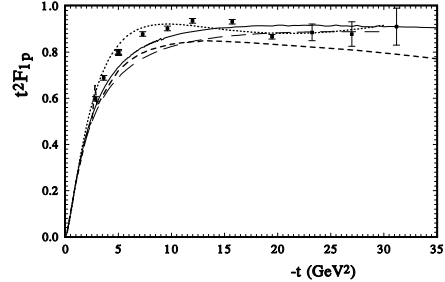} 
\caption{ Proton Dirac form factor multiplied by $t^2$
(the hard, dotted, long-dashed, and short-dashed lines correspond to our calculations with PDFs
 Al12, Stoller01, Rad04, and Kroll, respectively).
}
\label{Fig2}
\end{figure}

\section{GPDs and form factors of the nucleon }

   Further development of the model requires  careful analysis of the momentum transfer form of
   the GPDs and a properly chosen form the PDFs. In Ref. \cite{GPD-PRD14}, an analysis of
   more than 24 different
   PDFs was performed. We slightly complicated the form of the GPDs in comparison with Eq.(\ref{GPD0}),
   but it is the simplest one compared to other works (for example, Ref. \cite{Diehl-Kroll}):
\ba
{\cal{H}}^{u} (x,t) \  &=& q(x)^{u}_{nf} \   e^{2 a_{H}  \ \frac{(1-x)^{2+\epsilon_{u}}}{(x_{0}+x)^{m}}  \ t };  \\ \nonumber
{\cal{H_d}}^{d} (x,t) \  &=& q(x)^{d}_{nf} \   e^{2 a_{H} (1+\epsilon_{0}) (\frac{(1-x)^{1+\epsilon_{d}}}{(x_{0}+x)^{m}} ) \ t },
\label{t-GPDs-H}
\ea
\ba
{\cal{E}}^{u} (x,t) \  &=& q(x)^{u}_{fl} \   e^{2 a_{E}  \ \frac{(1-x)^{2+\epsilon_{u}}}{(x_{0}+x)^{m}}  \ t },  \\ \nonumber
{\cal{E_d}}^{d} (x,t) \  &=& q(x)^{d}_{fl} \   e^{2 a_{E}(1+\epsilon_{0}) (\frac{(1-x)^{1+\epsilon_{d}}}{(x_{0}+x)^{m}} ) \ t },
\label{t-GPDs-E}
\ea
 where $q(x)^{u,d}_{fl}=q(x)^{u,d}_{nf} (1.-x)^{z_{1},z_{2}}$.

 A complex analysis of the corresponding description of the electromagnetic form factors of the proton and neutron
    using the different  PDF sets  (24 cases) was carried out . These
   PDFs include the  leading-order, next-to-leading order, and the next-to-next-to-leading order
   determinations of the parton distribution functions. They used  different forms of the $x$ dependence of  the PDFs. 
   The analysis was carried out with different forms of the
   $t$ dependence of the GPDs. The minimum number of  free parameters was six and maximum was ten.

 To obtain the form factors,  we have to integrate  over $x$ in the whole range $0-1$.
 Hence, the form of the $x$ dependence of a PDF affects the form and size of the
  form factor.
 But the PDF sets are determined from the inelastic processes only in  some region of $x$, which is only
 approximated to $x=0$ and $x=1$.
   Some  PDFs  have a polynomial form of $x$ with a
     different power.  Some others have an exponential dependence on $x$.
  As a result, the behavior of the PDFs, when $x \rightarrow 0$ or $x \rightarrow 1$,  can  influence  the
    form of the calculated form factors.

 The sets of  experimental data are presented in Table I.
      The sets of  data have  various corrections and  different methods which take into account the
      systematic errors.  So we take into account only the statistical errors.
      On the basis of this analysis we calculated the electromagnetic form factors
      of the proton and neutron (using the isotopic symmetry).
      Then we carried out  the fit of these calculations and obtained
      the parameters of the electromagnetic [$G(t)$] and matter [$A(t)$] form factors.

    The results of the fitting procedure with different numbers of  free parameters
    for the whole set of  PDFs are presented in Fig. 1.
  We found that the best description was given by the PDFs from Refs. \cite{ABKM09,ABM12}.
   In this case, the increase in the    number of the free
   parameters leads to a small decrease in $\chi^2$. This means that the $x$ dependence of the PDFs
    corresponds sufficiently well to the $u$ and $d$ distributions in the nucleon  to reproduce the
   electromagnetic form factors. Note that these PDFs use a special power dependence on $x$.
    The most stable results (i.e., a minimum dependence on the number of free parameters
    with a minimum of $\chi^2$) are obtained with
    the PDFs    ABKM09 \cite{ABKM09} and ABM12 \cite{ABM12} (see Table II).

      The obtained form factors for the proton and neutron are shown in Fig. 2 and Fig. 3.
    The form factors practically  coincide for  both PDFs used.
    In Fig. 2, our results are compared with the
    other model calculations for $F_{1}(t)$.
    It should be noted that the experimental data for large $t$
    were obtained by the Rosenbluth method, and our calculations and the calculation
    in Ref. \cite{Diehl-Kroll}
    differ slightly  from the experimental data at large $t$, but they practically, coincide with each other.
      The ratio of $ \mu G_{E}/G_{M}$ for the proton and neutron cases is presented in Fig. 3.
      Our calculations reproduce the data obtained by the polarization method  quite well.

\begin{figure}
\includegraphics[width=.4\textwidth]{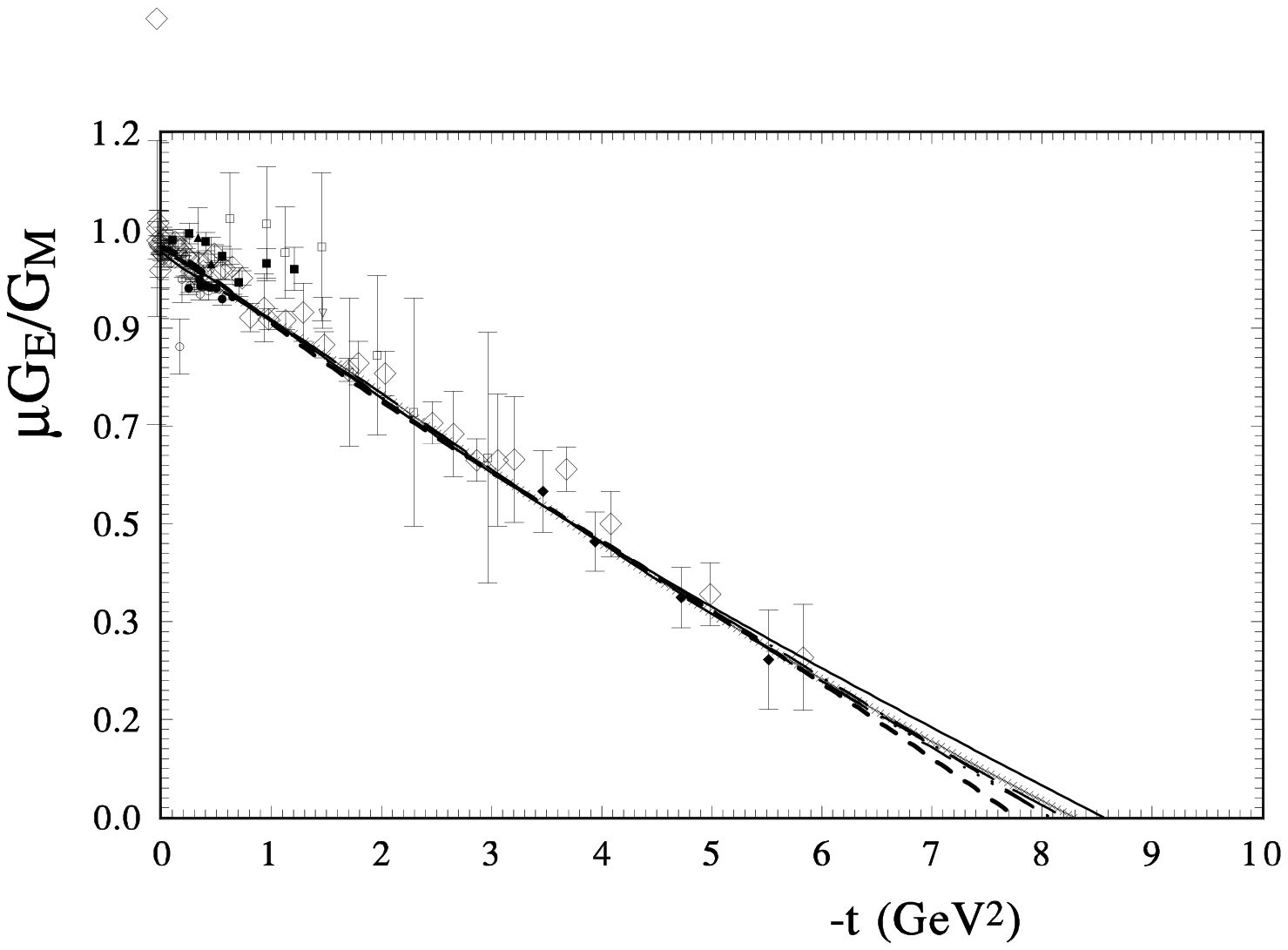} 
\includegraphics[width=.4\textwidth]{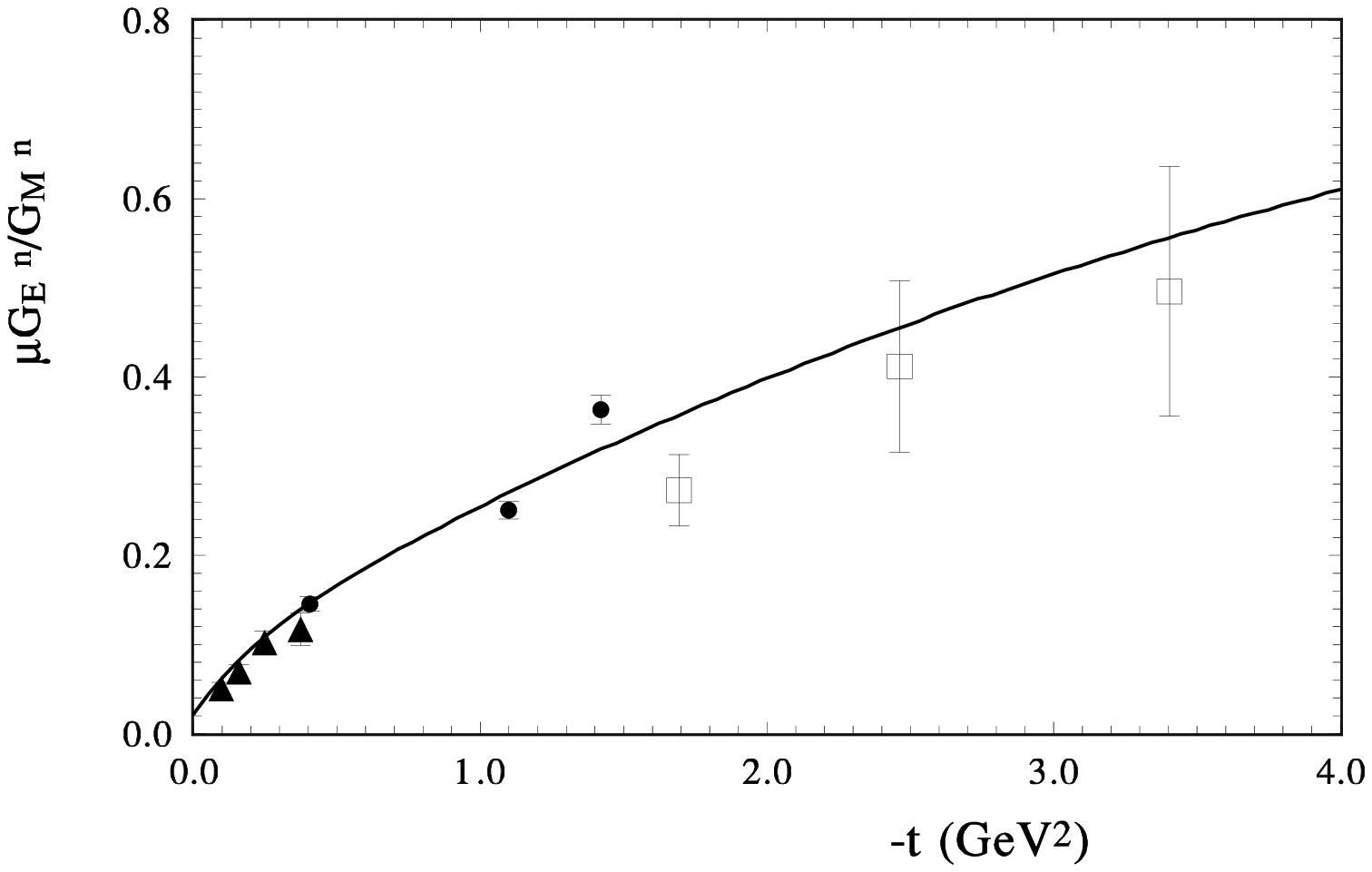} 
\caption{The model description of the ratio of the electromagnetic form  factors
   for the proton $\mu_{p} G^{p}_{E}/G^{p}_{M}$ and
   for the neutron $\mu_{n} G^{n}_{E}/G^{n}_{M}$.
  }
\label{Fig3}
\end{figure}

    On the basis of our GPDs with   
   the ABM12  PDFs \cite{ABM12},
 \ba
 q_{u}(x)=&&4.649903 x^{-0.288}(1.-x)^{3.637} \\	\nonumber
 &&x^{0.593 x-3.607 x^{2}+3.718 x^{3}},
\ea
\ba
     q_{d}(x)=&&3.424394 x^{-0.259}(1.-x)^{5.123} \\	\nonumber
    && x^{1.122 x-2.984 x^{2}},
\ea

    we calculated the hadron form factors
     using numerical integration,
 \ba
 F_{1}(t)&&= \int^{1}_{0}  dx
   \\	\nonumber
 &&[\frac{2}{3}q_{u}(x)e^{2 \alpha_{H} t (1.-x)^{2+\epsilon_{u}}/(x_{0}+x)^m} \\	 \nonumber
 &&  -\frac{1}{3} q_{d}(x)e^{ 2 \alpha_{H}  t (1.-x)^{1+\epsilon_{d}}/((x_{0}+x)^{m})} ]
\ea
   and then
    by fitting these integral results with the standard dipole form with some additional parameters
    for  $F_{1}(t)$,
      \begin{eqnarray}
   F_{1}(t)=\frac{4m_p-\mu t }{4m_p-t } \frac{1}{(1+q/a_{1}+q^{2}/a_{2}^2 +  q^3/a_{3}^3)^2 },
 \label{Gt}
 \end{eqnarray}
  The matter form factor 
 \ba
 A(t)&&=  \int^{1}_{0} x \ dx
 \\	\nonumber
 &&[ q_{u}(x)e^{2 \alpha_{H} t (1.-x)^{2+\epsilon_{u}}/(x_{0}+x)^m} \\	\nonumber
 &&  + q_{d}(x)e^{ 2 \alpha_{H}  t (1.-x)^{1+\epsilon_{d}}/((x_{0}+x)^{m})} ]
\ea
 is fitted   by the simple dipole form
      \begin{eqnarray}
   A(t)  =  \frac{\Lambda^4}{(\Lambda^2 -t)^2 }.
 \label{At}
 \end{eqnarray}
    The results of the integral calculations and the fitting procedure are shown in Fig. 4.
        Our description is valid up to a large momentum transfer
        with the following parameters: \\
        $a_{1}=16.7,  \ a_{2}^{2}=0.78, \ a_{3}^{3}=12.5$ and $\Lambda^2=1.6$.    \\
        These form factors will be used in our model of the proton-proton and proton-antiproton elastic scattering.

\begin{figure}
\includegraphics[width=.4\textwidth]{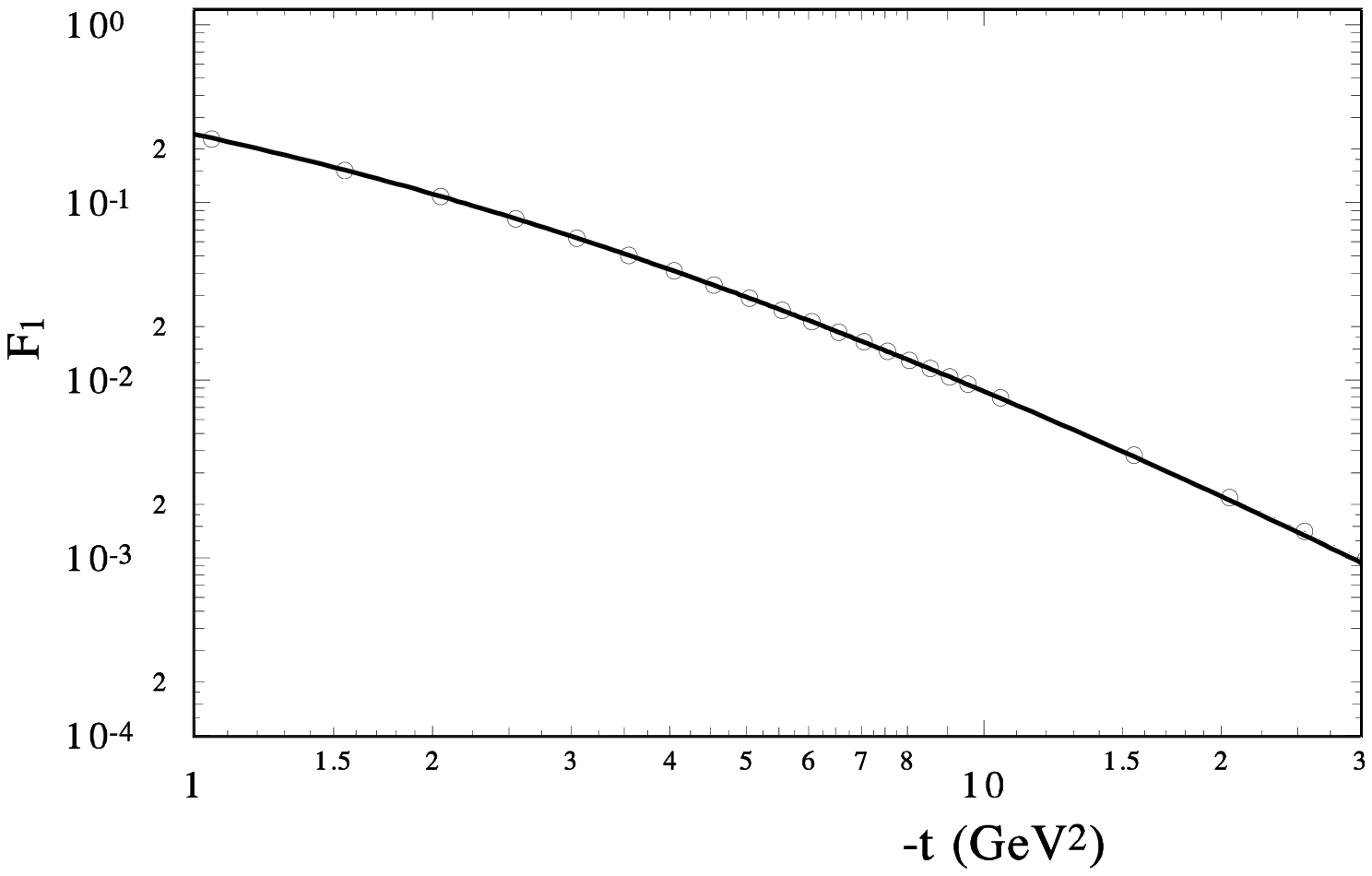} 
\includegraphics[width=.4\textwidth]{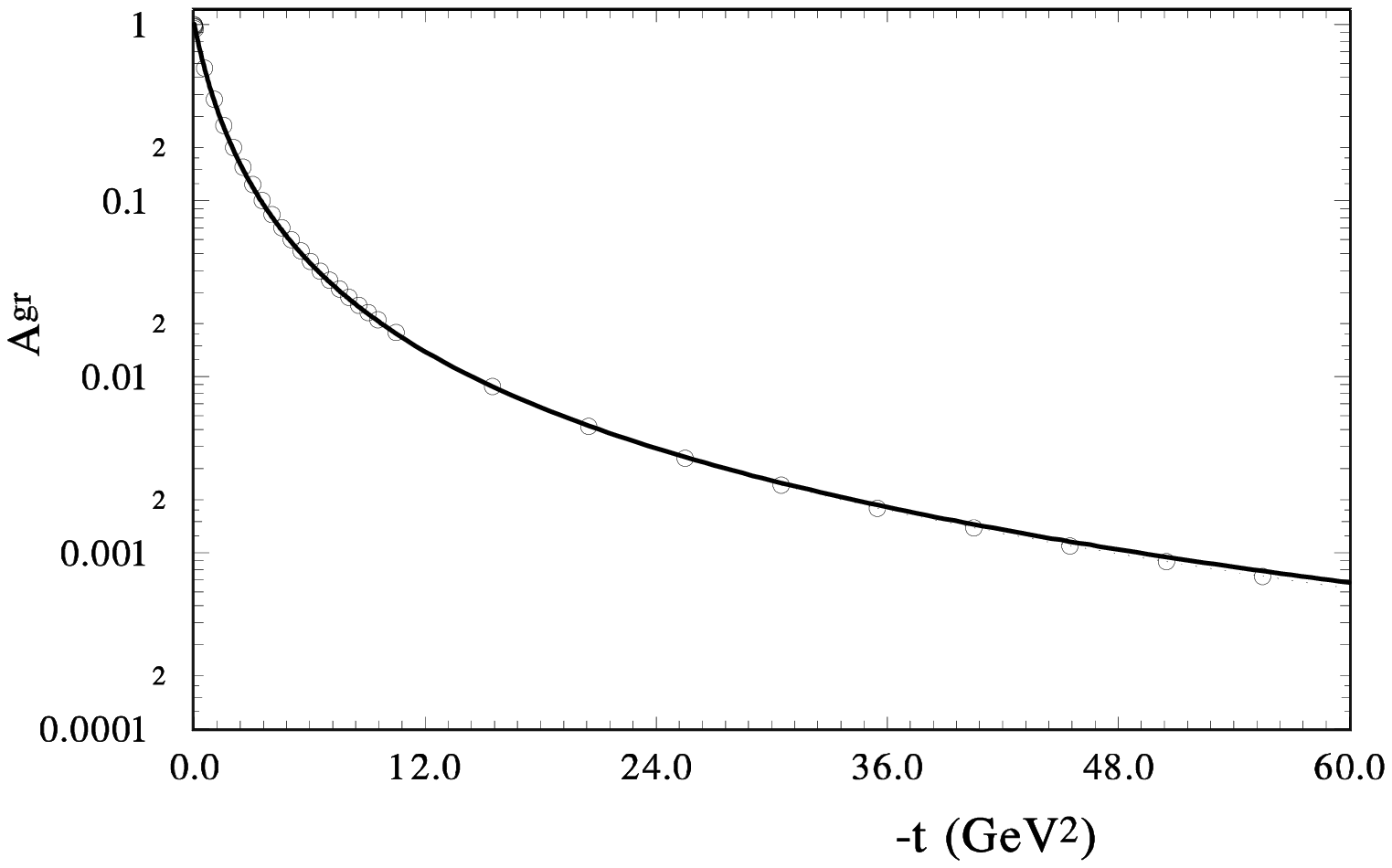} 
\caption{ The fit of the form factors of the proton:
(a) (top), the electromagnetic form factor $G(t)$ [Eq.(\ref{Gt})] and
and (bottom) the matter form factor $A(t)$ [Eq.(\ref{At})].
 The circles are the moments of the GPDs (shown only every tenth point).
  }
\label{Fig4}
\end{figure}

\section{Extension of the HEGS model}

    The  obtained form factors differ slightly from those used in our previous work \cite{HEGS-JEP12}.
      Hence, we have to make a new fit of high-energy data, including now the new data of the TOTEM
      Collaboration \cite{TOTEM-1395,TOTEM-8nexp}.
      As was noted in our previous work, the model also describes low-energy data qualitatively.
      Now we include in our fitting procedure additional experimental data
      on the $pp$ and $p\bar{p}$ elastic scattering up to $8$ TeV $\geq \sqrt{s} \geq 9.8 $ GeV.
      As a result, the amount of  experimental data increases by a factor of $3.5$ (from $980$ to $3416$).
      This gives us many experimental high-precision data points at small momentum transfer, including
      the Coulomb-hadron interference region
      where the experimental errors are remarkably small.
      Hence, we can check  our model construction
      where the real part is determined only by the complex representation of $\hat{s}=s/s_{0} exp(-i \pi /2)$.
  We do not include the data on  the total cross sections
        $\sigma_{\rm tot}(s)$  and $\rho(s)$, as their values were obtained from the differential
        cross sections, especially in the Coulomb-hadron interference region.
         Including such data decreases $\chi^2$, but it  would be double counting in our opinion.
          We also do not include the interpolated
         and extrapolated data of Amaldi \cite{Amaldi-166}, and only include their original experimental data.

      As in the old version of the model, we  take into account only the statistical errors in the standard
      fitting procedure.
       The systematic errors are taken into account by the additional normalization coefficient
       which is the same for  every row of the experimental data.
       It essentially decreases the space of the possible form of the scattering amplitude.
        Of course, it is necessary to control the sizes of the normalization coefficients
         so that they do not introduce an additional energy dependence.
      As we will see later (Tables III and IV), the distribution of the coefficients has the
        correct statistical properties
      and does not lead to a visible additional energy dependence.

      Such a simple form of the scattering amplitude in the huge region of energy requires
      careful determination of the slope of the scattering amplitude.
  As was noted in Ref. \cite{Rev-LHC}),
  analytic $S$-matrix theory, perturbative quantum chromodynamics,
and the data require that Regge trajectories be nonlinear complex
functions \cite{nonlin1,Jenk-nl}. 
  The Pomeron trajectory has threshold singularities, the lowest one
being due to the two-pion exchange, required by the $t$-channel
unitarity Ref. \cite{Gribov-Sl}.
This threshold singularity appears in different forms in various
models (see \cite{Rev-LHC}).

  In the present model, 
     a  small additional term is introduced into the slope which reflects some possible small nonlinear
        properties of the intercept.
     As a result, the slope is taken in the form
   \begin{eqnarray}
   B(s,t) \ =  (\alpha_{1} + k q e^{-k q^2 Ln(\hat{s} \ t)} ) Ln(\hat{s}) .
   \label{B0}
\end{eqnarray}
   This form leads to the standard form of the slope as $t \rightarrow 0$ and $t \rightarrow \infty$.
   Note that our additional term at large energies has a similar form as an additional term to the slope
   coming from $\pi$ loop examined in Ref. \cite{Gribov-Sl} and recently in Ref. \cite{Khoze-Sl}.
   The basic Born amplitudes were taken in the old form,
     [Eqs. (5)  and (\ref{FB-ab})]
    with  fixed $\alpha_{1}=0.24$ GeV$^{-2}$ and  $\Delta=0.11 $.
   Taking into account the Mandelstam region of the analyticity scattering amplitude
   for the $2 \rightarrow 2 $ scattering process
   $s+u+t = 4 m_{p}^2$   we take $ s_{0}=4 m_{p}^{2}$ where
     $m_{p}$ is the mass of the proton.

      Then, as we intend to describe  sufficiently low energies,  possible Odderon contributions
      were taken into account: 
  \begin{eqnarray}
 F_{\rm odd}(s,t) \ =  \pm   \ h_{\rm odd} \ A^{2}(t) \ F_{b}(s,t),
\end{eqnarray}
 where $h_{\rm odd} = i h_{3} t/(1-r_{0}^{2} t) $.

   Just as we supposed in the previous variant of the HEGS model that  $F_{b}(s,t)$ 
    corresponds to the
   cross-even part of the three gluon exchange, our Odderon contribution is also connected
   with the matter form factor $A(t)$.
   Our ansatz for the Odderon slightly differs from the cross-even part by some kinematic  
    function.
 The form of the Odderon working in  all $t$
    has the same behavior as the cross-even part at larger momentum transfer, of course,
   with different signs for proton-proton and proton-antiproton reactions.
   It has a large preasymptotic part, and as a result, such a preasymptotic part of the cross-even part
   is practically not felt.
      Hence, the Born term of the elastic hadron amplitude can now be written as
%
  \begin{eqnarray}
 F_{h}^{Born}(s,t)=&&h_1 \ F_{1}^{2}(t) \ F_{a}(s,t) \ (1+r_1/\hat{s}^{0.5})   \\ \nonumber
     &&+  h_{2} \  A^{2}(t) \ F_{b}(s,t) \     \\
     &&\pm h_{odd} \  A^{2}(t)F_{b}(s,t)\ (1+r_2/\hat{s}^{0.5}),  \nonumber
    \label{FB}
\end{eqnarray}
  where $F_{a}(s,t)$ and $F_{b}(s,t)$  are the same as in the previous variant of the model [see Eq. (\ref{FB-ab})].


   The analysis of the hard Pomeron contribution in the framework of the  model    \cite{NP-HP}
    shows that such a contribution is not felt. For the most part, the fitting procedure requires a negative  additional hard Pomeron contribution.
     We repeat the analysis of  \cite{NP-HP} in the present model
    and obtain practically the same results.
    Hence, we do not include the hard Pomeron in the model.

      At large $t$  our model calculations are extended up to $-t=15 $ GeV$^2$.
  We added a small contribution from the energy-independent  part
  of the spin-flip amplitude in a form similar to that  proposed    in Ref. \cite{Kuraev-SF}.
  \begin{eqnarray}
  F_{sf}(s,t) \ =  h_{sf} q^3 F_{1}^{2}(t) e^{-B_{sf} q^{2}}.
  \end{eqnarray}
  It has two additional free parameters.  Of course, at  lower energy we need to take into account
  the energy-dependent parts of the spin-flip amplitudes.
  However, this requires including additional
  polarization data in our examination which essentially complicates the picture.  This is  beyond the scope of this paper.
  Such a contribution  can be made in future works.
   The model is very simple from the viewpoint of the number of fitting parameters and functions.
  There are no  artificial functions or any cuts which bound the separate
  parts of the amplitude by some region of momentum transfer.

\begin{figure}
  \includegraphics[width=.4\textwidth]{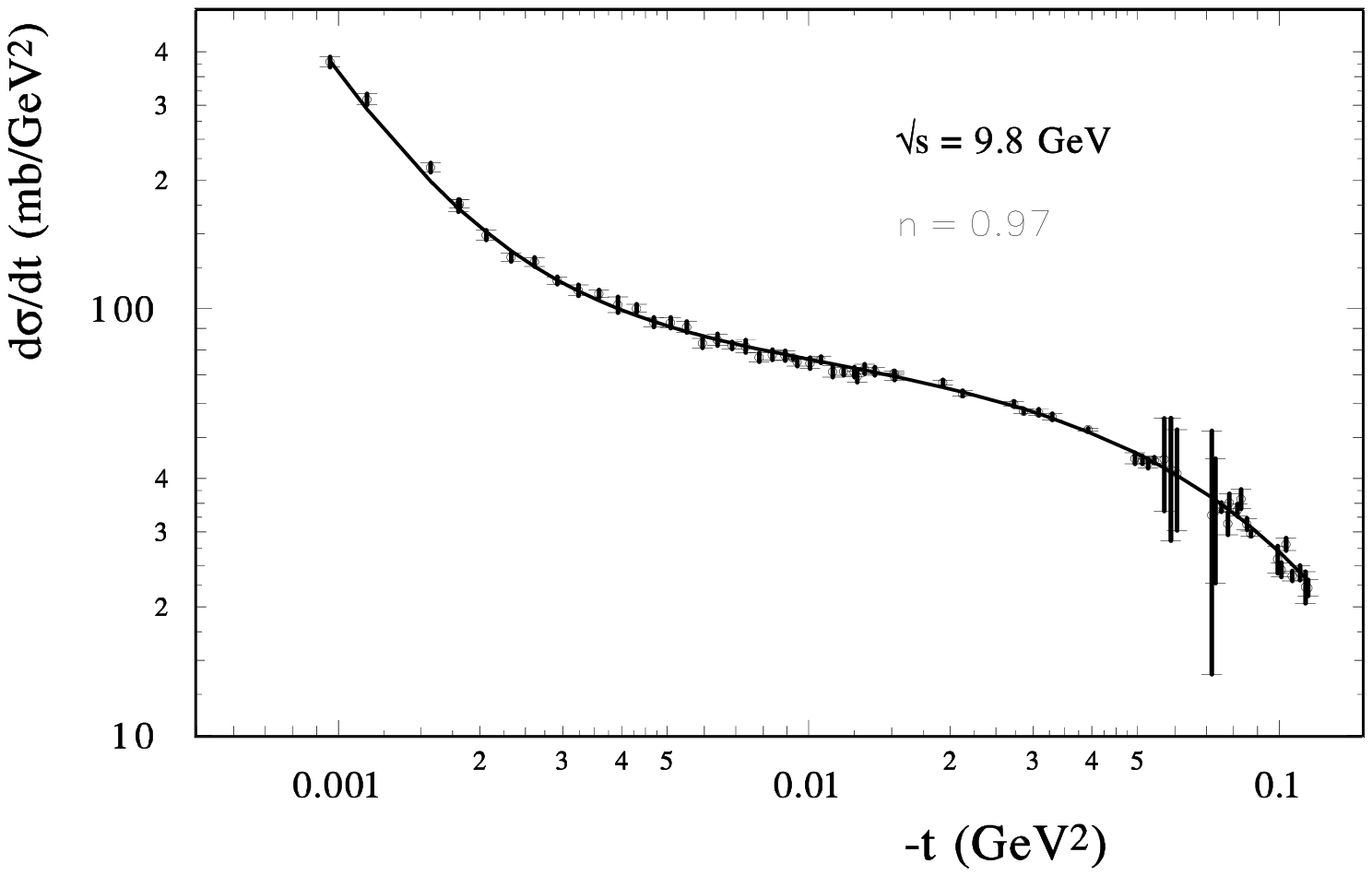}
      \includegraphics[width=.4\textwidth]{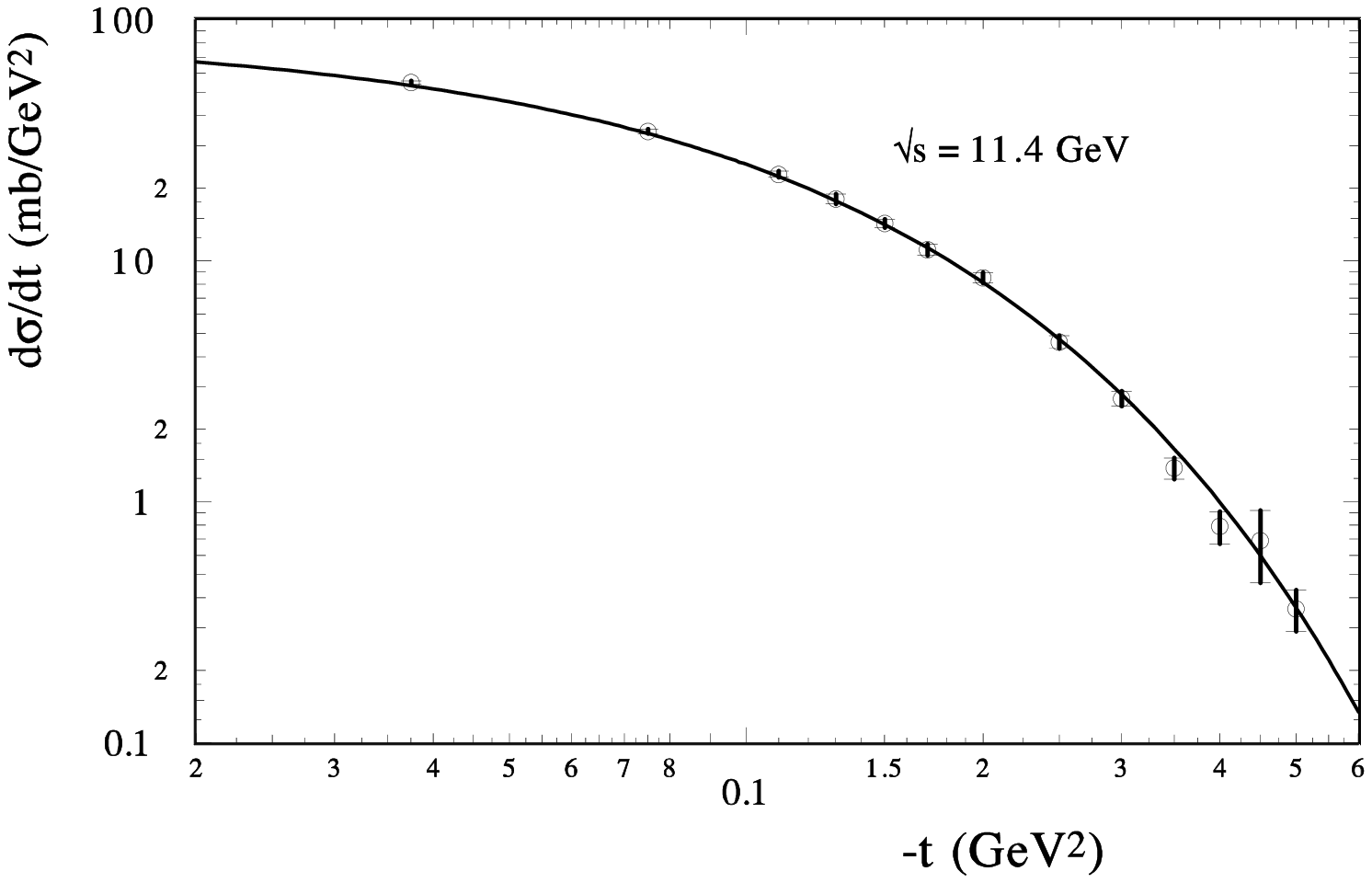}
  \caption{
  $d\sigma/dt$ for $pp$ (top) at $\sqrt{s}=9.8$ GeV
    and $p\bar{p}$ (bottom) at $\sqrt{s}=11.3$ GeV.
   }
 \label{Fig5}
\end{figure}

\section{Analysis and Results}

 We included $3416$ experimental points were included in our analysis
 in the energy region   $9.8$ GeV $\leq \sqrt{s} \leq 8. $ TeV
 and in the region of momentum transfer $0.000375 \leq |t| \leq 15 $ GeV$^2$.
 The experimental data of the proton-proton and proton-antiproton elastic scattering are included
 in 92 separate rows of 32 experiments \cite{data-Sp}, including recent data from the TOTEM Collaboration
 at $\sqrt{s}=8$ TeV  \cite{TOTEM-8nexp}.
  The whole Coulomb-hadron interference region,
  where the experimental errors are remarkably small,
    was included in our examination of the experimental data (see Tables III and IV).

   In the fitting procedure by FUMILIM program \cite{Sitnik} we calculated the minimum in  $\sum_{i=1}^{N} \chi_{i}^2$
   related to the statistical errors $\sigma_{i}^{2}$. The systematic errors
   are taken into account by the additional normalization coefficient $n_{k}$
   for the $k$ series (the experiment) of experimental data,
\begin{eqnarray}
\chi^2= \sum_{i=1}^{N} \frac{n_{k} \ E_{i}^{k}(s,t) \ - \  T_{i}(s,t)}{\sigma_{i}^2(s,t)}, \label{fit}
 \end{eqnarray}
 where  $T_{i}(s,t)$ are the theory predictions, including  the hadronic and electromagnetic parts
  of the scattering amplitude, and $n_{k} E_{i}(s,t)$ are the  data points allowing a shift by the systematical error of the $k$ experiment (see, for example, Refs.  \cite{pdf-system1,pdf-system2}.

In the region of  small momentum transfer the systematic errors
 are on the order of $2-5 \%$.
 For the most part, the additional normalization is in the region $0.95-1.05$.
    At large momentum transfer the order of the systematical errors is
    $10-20\% $.
    In this case, the additional normalization is situated in the region $0.8-1.2$.
   Of course, if one sums the systematic and
    statistical errors,  $\sum \chi^2/N $ decreases
    but it will give some additional space for  changing the form of the scattering amplitude.

\begin{figure}
\vspace{-0.7cm}
    \includegraphics[width=.45\textwidth]{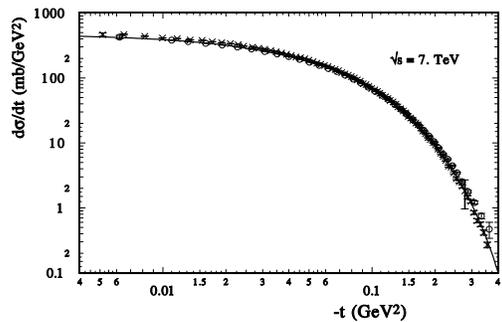}
\vspace{-0.6cm}
         \includegraphics[width=.45\textwidth]{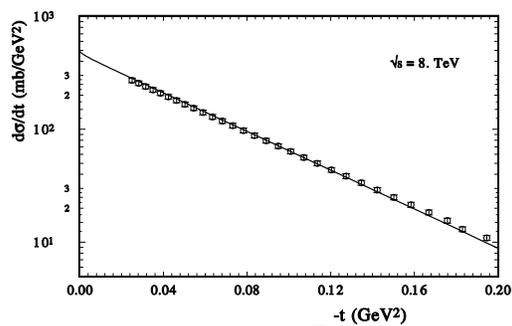}
  \caption{
  $d\sigma/dt$ for $pp$ (top) at $\sqrt{s}=7$ TeV
  (crosses and circles are  TOTEM \cite{TOTEM-11} and ATLAS \cite{ATLAS-14} data, respectively)
 and  (bottom) at $\sqrt{s}=8$ TeV (circles are TOTEM data \cite{TOTEM-8nexp}]
   }
 \label{Fig6}
\end{figure}

     Our complete fit of 3416 experimental data points in the energy range
      $9.8 \leq \sqrt{s} \leq 8000 \ $ GeV
      and the region  of  momentum transfer
      $0.000375 \leq \ -t \ \leq 14.75 \ $GeV$^2$ gives
  $ \sum_{i=1}^{N} \chi_{i}^{2}/N=1.28$,
with the parameters
$h_1=3.67; \ h_2=1.39; \ h_{odd}=0.76; \
 k_0=0.16; \  r_{0}^{2}=3.82$, and the low-energy parameters
  $h_{sf} = 0.05; \ r_1=53.7; \ r_2 = 4.45$.

           Obviously, for such a huge energy region we have a very small
   number of free parameters.
  We also note  the good description of the CNI region of momentum transfer
  in a very wide energy region
   (approximately 3 orders of magnitude) with the same slope of the scattering amplitude.
 The differential cross sections 
 of the proton-proton and proton-antiproton elastic scattering
  at small  momentum transfer 
  are presented in  Fig. 5 at   $\sqrt{s}= 9.8 $ GeV   for $pp$ scattering,
   and $\sqrt{s}= 11 $ GeV for  $p\bar{p}$ elastic scattering,
   and in Fig. 6 at $\sqrt{s}= 7. $ TeV and $\sqrt{s}= 8. $ TeV for $pp$ scattering.
   The model quantitatively reproduces the differential cross sections in the whole examined energy region
 in spite of the fact that the size of the slope  is essentially changing in this region
 [due to the standard Regge behavior $log(\hat{s}$]
  and the real part of the scattering amplitude has  different behaviors for
   $pp$ and $p\bar{p}$.

     The results for the whole energy region for small momentum transfer
     are presented in  Table III for the proton-proton  elastic scattering
     and in  Table IV for the proton-antiproton elastic scattering.
     We can see that the $\chi^2$ values are suitable.
     We note that they are  especially small for the high-precision FNAL-JINR data
     which reach a very small size of the momentum transfer (up to $-t=0.00037$ GeV$^2$   \cite{Kuznetzov}
     at energies $\sqrt{s} = 13.4, \ 19.4, \ 23.4$, and $27.4$ GeV.
     The additional normalization coefficients do not show an energy dependence
     and are distributed only  statistically. We also include the high-precision
      non-normalized data of the UA4/2 Collaboration at $\sqrt{s}=541 $ GeV
      which reach a very small momentum transfer $-t_{\rm min}=0.000875$ GeV$^2$.

\begin{table}
 \caption{The proton-proton elastic scattering at small $t$}
\label{Table-1}
\begin{center}
\begin{tabular}{|c|c|c|c|c|c|c|} \hline
$\sqrt{s}$,  & $t_{
\rm min}$, & $t_{\rm max}$, & N & $\sum_{N} \chi^{2}$  & $\sum_{N} \chi^{2}/N$ &$n_{k}$ \\ 
 GeV     &   GeV        &     GeV           &        &     & & (norm.)  \\ \hline
 9.0   &0.00193    & 0.04328 & 19 & 14.4   &  0.72   &  1.041  \\
 9.3   &0.01268    & 0.1147  & 28 & 21     &  0.75   &  1.019   \\
 9.8   &0.00115     & 0.115  & 64 & 87.6   &  1.35   &  1.013   \\
 9.8   &0.0026     & 0.12    & 23 & 31.0   &  1.37   &  1.074   \\
 9.9   &0.00063    & 0.0306  & 73 & 81.1   &  1.11   &  1.014   \\
 10.6  &0.00079    & 0.01529 & 45 & 68.0   &  1.39   &  1.026   \\
 12.3  &0.00066    & 0.02928 & 58 & 46.9   &  0.81   &  1.018  \\
13.76  &0.0023     & 0.0388  & 73 & 84.9   &  1.16   &  1.023   \\
13.76  &0.035      & 0.095   & 7  & 2.5    &  0.36   &  1.029   \\
16.83  &0.0022     & 0.0392  & 68 & 76.9   &  1.13   &  1.006   \\
 19.42 &0.00066    & 0.0315  & 69 & 79.5   &  1.15   &  0.996 \\
 19.42 &0.035      & 0.095   & 7  & 12.2   &  1.74   &  1.008  \\
 19.42 &0.0206     & 0.12    & 42 & 19.9   &  0.47   &  1.038  \\   
  21.7 &0.022      & 0.039   & 64 & 50.1   &  0.78   &  0.996   \\
 22.2  &0.0005     & 0.02978 & 64 & 55.6   &  0.87   &  1.007   \\
 23.5  &0.00037    & 0.0102  & 30 & 58.5   &  1.95   &  1.008   \\
 23.8  &0.0022     & 0.0388  & 60 & 69.1   &  1.15   &  1.001  \\
 23.9  &0.00066    & 0.0316  & 66 & 76.5   &  1.16   &  0.988  \\
 27.4  &0.00047    & 0.02579 & 61 & 66.1   &  1.08   &  0.987  \\
 30.6  &0.016      & 0.11    & 48 & 53.1   &  1.10   &  1.005  \\
 30.8  &0.0005     & 0.0176  & 31 & 75.7   &  2.36   &  1.009 \\
 44.7  &0.00099    & 0.01856 & 40 & 51.    &  1.16   &  1.004   \\
 52.8  &0.00107    & 0.05546 & 35 & 53.2   &  1.52   &  1.016   \\
 62.3  &0.00543    & 0.05122 & 23  & 31.7  &    1.38 &  1.005   \\
 7000. &0.00515    & 0.356   & 84  &173.4  &    2.04  &  0.943 \\
 7000. &0.006      & 0.36    & 40  & 31.4    &    0.77  &  1.0  \\
 8000. &0.028      & 0.195   & 30  & 20    &    0.7  &  0.9  \\
 \hline
\end{tabular}
\end{center}
  \end{table}

\begin{figure}
\vspace{-0.7cm}
      \includegraphics[width=.45\textwidth]{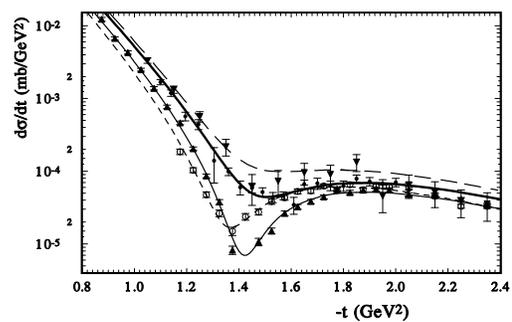}
 \vspace{-0.4cm}
  \caption{The form and energy dependence of the diffraction minimum
  at low energies (long-dashed line: $\sqrt{s}=13.4$ GeV;
  thick hard line: $\sqrt{s}=18.4$ GeV; hard line: $\sqrt{s}=30.4$ GeV;
  dashed line: $\sqrt{s}=44.7$ GeV).
   }
\label{Fig7}
\end{figure}
\begin{figure}
      \includegraphics[width=.4\textwidth]{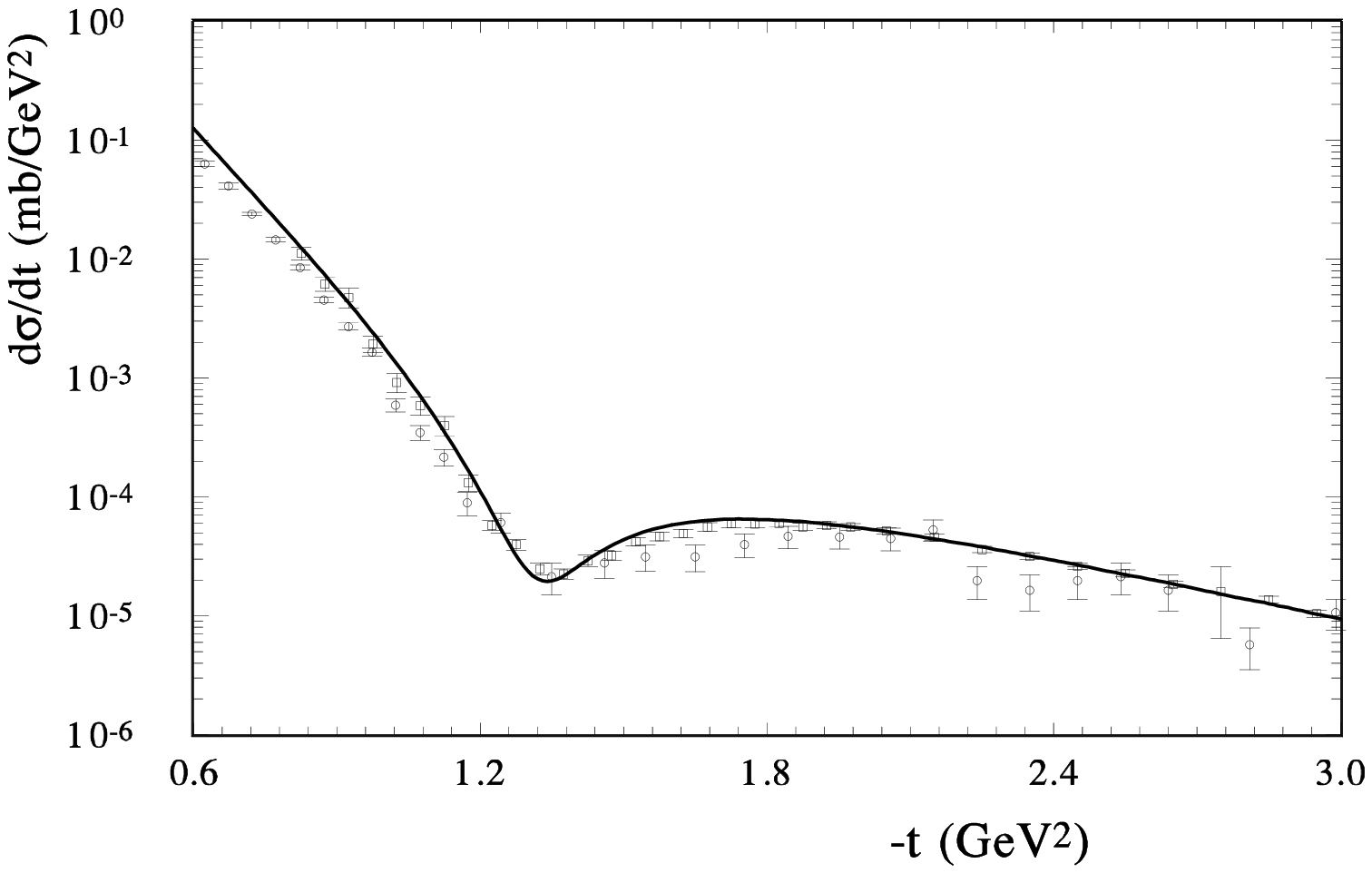}
    \includegraphics[width=.4\textwidth]{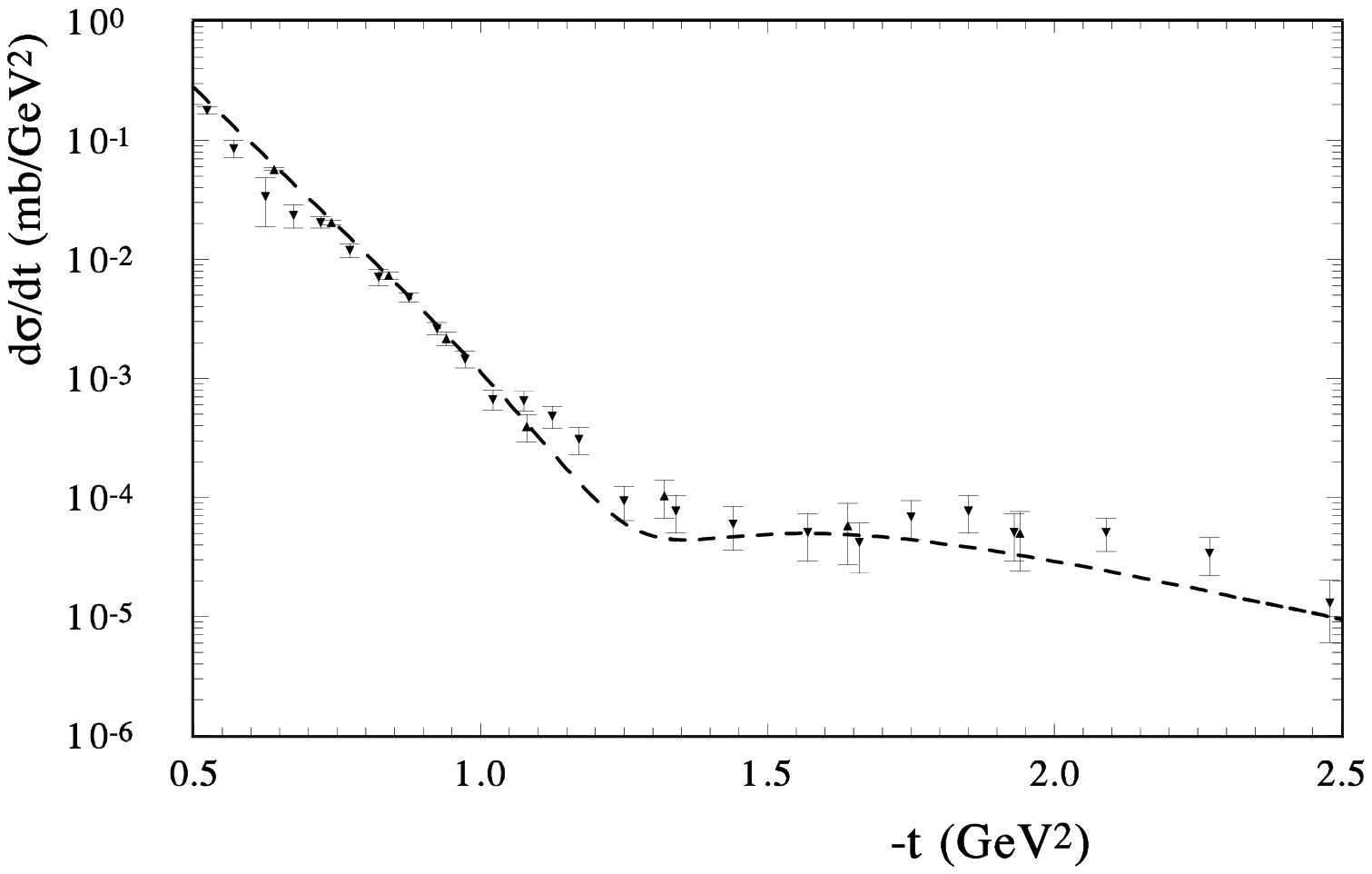}
            \includegraphics[width=.4\textwidth]{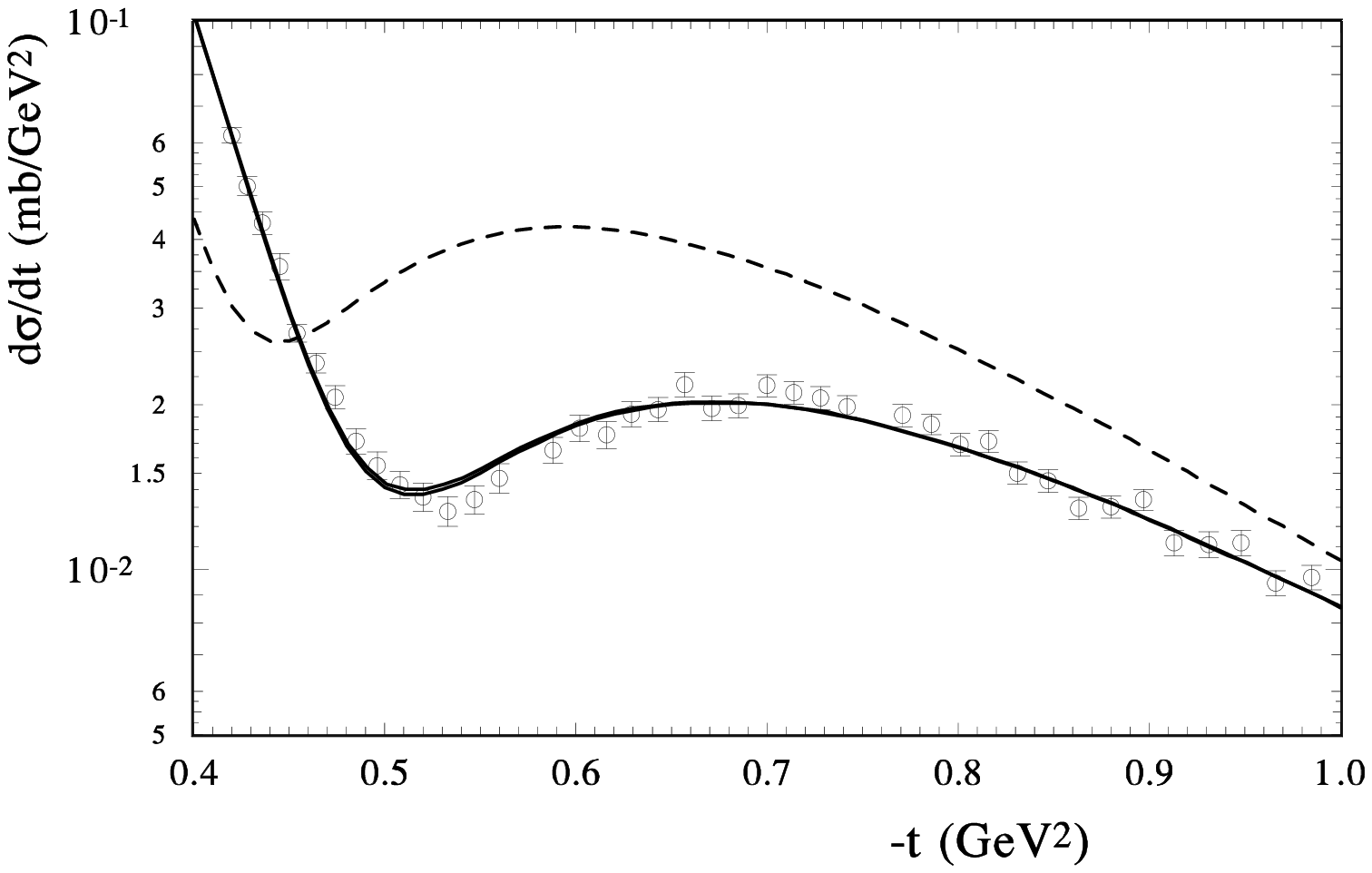}
  \caption{
  $d\sigma/dt$ for $pp$  at $\sqrt{s}=52.8$ GeV  (top),
    and  $p\bar{p}$  at $\sqrt{s}=52.8$ GeV (midle), and
      $pp$  at $\sqrt{s}=7$ TeV (bottom) [the dashed line (bottom)
       is the predictions at $\sqrt{s}=14$ TeV].
   }
 \label{Fig8}
\end{figure}
\begin{table}
 \caption{The proton-antiproton elastic scattering at small $t$}
\label{Table-3}
\begin{center}
\begin{tabular}{|c|c|c|c|c|c|c|} \hline
$\sqrt{s}$,  & $t_{\rm min}$,  & $t_{
\rm max}$,  & N & $\sum_{N} \chi^{2}$  & $\sum_{N} \chi^{2}/N$ &$n_{k}$ \\     \hline
   GeV & GeV & GeV &           &               &             &    (norm.)             \\ 
 11.54  &0.0375    & 0. 5   & 13 & 11.5  &  0.88 &  0.983   \\
 13.76  &0.035    & 0.095   & 7  & 7.4   &  1.06 &  0.966   \\
 19.42  &0.035    & 0.095   & 7  & 7.3   &  1.05  & 1.220 \\
  30.4  &0.00067  & 0.01561 & 28 & 28.8  &  1.03  & 0.974  \\
 52.6  &0.00097   & 0.03866 & 28 & 24.5  &  0.875 & 0.987   \\
 52.8  &0.0109    & 0.0479  & 43 & 49.9  &  1.16  & 0.933\\
 62.3  &0.00632   & 0.03821 & 43 & 55.8  &  1.3   & 0.996   \\
 541.  &0.000875  & 0.11875 & 99 & 164.7 &  1.65  &  unnorm.  \\
 546.  &0.00225   & 0.03475 & 66 & 83.7  &  1.25  &  1.004  \\
 546.6 &0.026     & 0.078   & 14 & 13.86 &  1.0   &  1.002  \\
 1800. &0.0339    & 0.285   & 28 & 28.8  &  1.03  &  1.024  \\
   &         &              &                &   &        \\
 \hline
\end{tabular}
\end{center}
  \end{table}

     The real part of the scattering amplitude significantly influences  the size
     and form of the differential cross sections
     in the Coulomb-hadron interference region  \cite{Sel-UA42,TOTEMrho-NPhis14}.
       The second Reggeons also have a large slope for the imaginary part  and  hardly
       change the slope of the differential cross sections.
       A suitable description of  both $pp$ and $p\bar{p}$ experimental data in this region
       supports the determination of the real part of the scattering amplitude chosen in the model.
       It should be  noted that  possible contributions of the second Reggeons will
       essentially change the form and size of the real part of the scattering amplitude.
       The results presented in  Tables III and IV show that up to such a low energy we do not
       feel the essential contributions of the second Reggeons.
       This especially  concerns the possible contribution of the $f_{0}$ meson.
       In some models it has an intercept essentially above $0.5$ and
       its contribution influences the differential cross sections
       and $\sigma_{\rm tot}(s)$ and $\rho(s,t)$
        in the ISR energy region. Our results practically exclude such Reggeons with
        an intercept  above $0.5$.

   The form and energy dependence of the diffraction minimum  are very sensitive
   to the different parts of the scattering amplitude. The change of the sign of the imaginary part
   of the scattering amplitude determines the position of the minimum and its movement
    with a change in the energy.
   The real part of the scattering amplitude determines the size of the dip. Hence, it depends
   heavily on the Odderon contribution. The spin-flip amplitude gives the contribution in the differential cross  sections additively.
   So the measurement of the form and energy dependence of the diffraction minimum
   with high precision is an important task for future experiments.
     In Fig. 7, the description of the diffraction minimum in our model is shown for low energies.
     The HEGS model sufficiently reproduces the energy dependence and  form of the diffraction dip.
     In this energy region the diffraction minimum reaches the sharpest dip at  $\sqrt{s}=30 $~GeV.
     Note that at this energy the value of $\rho(s,t=0)$ also changes its sign in the proton-proton
     scattering.
     The $p\bar{p}$ cross sections in the model are obtained by the $s \rightarrow u$
       crossing without changing the model parameters.
      And  for the proton-antiproton scattering
       the same situation  with correlations between the sizes of  $\rho(s,t=0)$ and $\rho(s,t_{\rm min})$
       takes  place at low energy (approximately  at $p_{L}= 50 $ GeV).
     Such a correlation was noted in Ref. \cite{Sel-YF87}.

The model reproduces   $d\sigma/dt$ at very small and large $t$ and provides a qualitative description of the dip region
at $-t \approx 1.4 $~GeV$^2 $, for $\sqrt{s}=53 $~GeV and for $\sqrt{s}=62.1 $~GeV
for the proton-proton and proton-antiproton elastic scattering (see  top and middle panels of Fig.8).
The diffraction minimum at $\sqrt{s}=7 $~TeV
 is reproduced sufficiently well too (see Fig.8c).

In Fig.~9, the description of the differential cross sections of the elastic scattering  $pp$ 
  at large $t$ and different values of $s$ is presented.
   It is to be  noted that the calculation of our  integrals with complex oscillation functions
   at large momentum transfer is a difficult task    and requires high-precision of the calculations.
   In any case,
   we obtain  a quantitatively good description of the differential cross sections
   at  large $t$.
   In this region of $t$, the contribution of the spin-flip amplitude is felt.
   We take into account only the asymptotic part of this amplitude with the simplest
   and  energy-independent forms.
   Although it has a small size, its constant is determined sufficiently well,
     $h_{sf} =0.06 \pm 0.004$.

  In Fig. 10, the model calculations for the differential cross sections are shown for the LHC energies.
  Obviously, in the model the difference in the behavior of the differential cross sections between
  $\sqrt{s}=7 $ and  $\sqrt{s}=14 $ TeV is not large.
  For the most part, it is reflected in the movement of the position of
  the diffraction minimum to low momentum transfer and the increase of  the sizes of the differential cross sections  in the minimum and second diffraction maximum.

\begin{figure}
  \includegraphics[width=.4\textwidth]{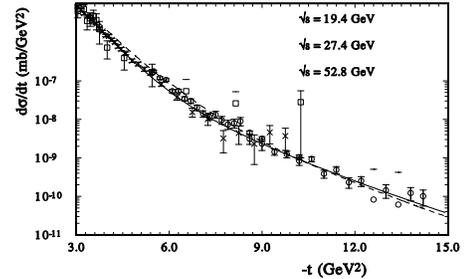}
  \caption{ $d\sigma/dt$ at large $t$  for $pp$  at $\sqrt{s}=19.4$ GeV (hard line),
           $\sqrt{s}=27.4$ GeV, and $\sqrt{s}=52.8$ GeV (dashed line).
         The   squares, circles, crossing are  the experimental data for each case, respectively. 
  }
 \label{Fig9}
\end{figure}

\begin{figure}
    \includegraphics[width=.4\textwidth]{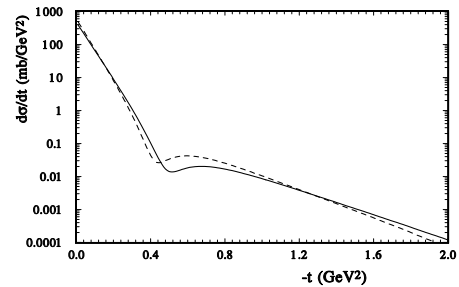}
  \caption{  
     The model predictions    of  $d\sigma/dt$ at $\sqrt{s}=7$ (hard line) and $\sqrt{s}=14$ TeV
     (dashed line).
  }
 \label{Fig10}
\end{figure}
\begin{figure}
  \includegraphics[width=.4\textwidth]{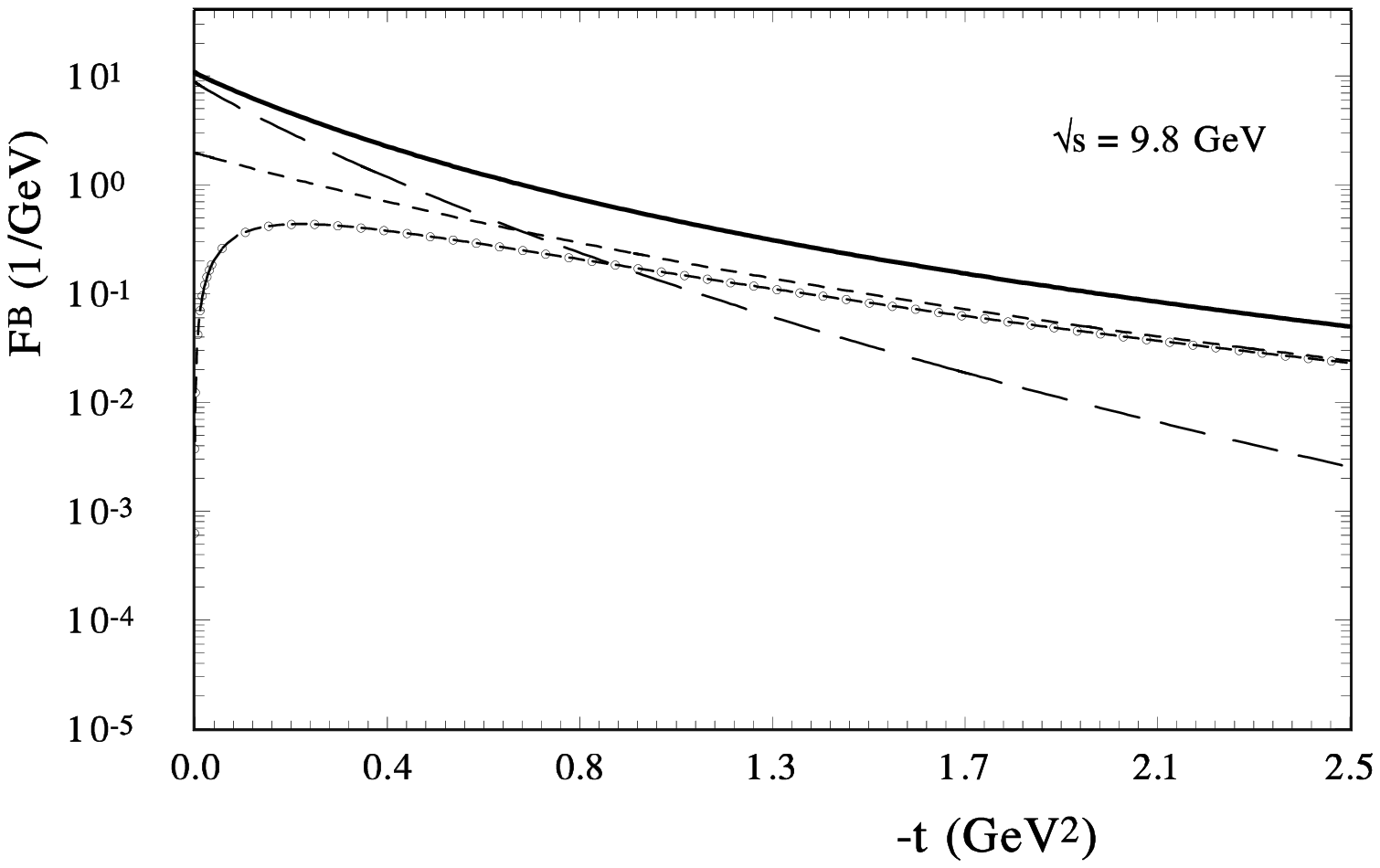}
      \includegraphics[width=.4\textwidth]{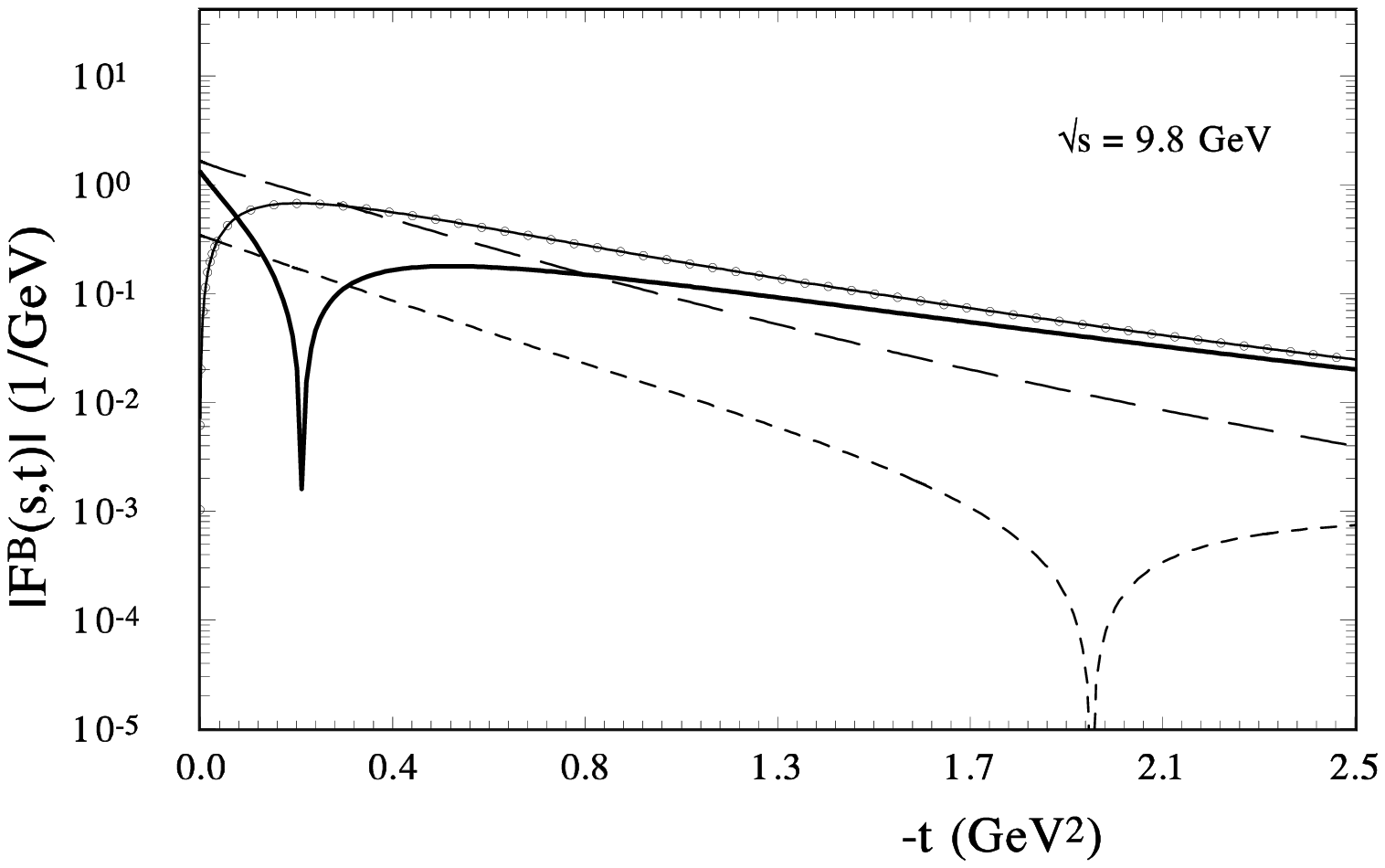}
  \caption{The magnitude of the Born parts of  the $pp$ elastic scattering amplitudes
  at $\sqrt{s}=9.8$ GeV.
   (Top) The imaginary parts:  the sum of all parts (hard line),
    the contribution of  $P_{2g}$ (long-dashed line),
    the contribution of the cross-even $P_{3g}$  (dashed line),
   and the contribution of the cross-odd part of $P_{3g}$ (line with points).
  (Bottom) The same for the real parts of the Born amplitude.
   }
 \label{Fig11}
\end{figure}

\section{The structure of the scattering amplitude}

   In the model, only the Born terms of the scattering amplitude are determined.
   The separate terms and the full Born scattering amplitude have a simple form and
   their imaginary parts do not have any oscillating behavior at  small momentum transfer.
   In Fig. 11 and Fig. 12, the parts of the Born terms of the scattering amplitude
    are presented at    $\sqrt{s}=9.8$ GeV and  $\sqrt{s}=7.$ TeV.
   At small momentum transfer the imaginary part of the Pomeron $P_{2g}$ dominates at both energies.
    At small energy, the cross-even and cross-odd imaginary parts of $P_{3g}$ are equal to  $P_{2g}$
    in the region of momentum transfer $0.7-0.9$ GeV$^2$, and then they dominate. Both parts of $P_{3g}$
    have the same size at $-t > 1.$ GeV$^2$. But the real part of the cross-odd term of $P_{3g}$
    dominates  at $-t> 0.4 $ GeV$^2$ and
    determines the real part of the full Born term of the scattering amplitude.
     As it has a different sign, compared to the real part of $P_{2g}$,
      the full real part of the Born term
     changes the sign at $-t=0.3$ GeV$^2$.

     At $\sqrt{s}=7$ TeV the picture is different. In this case, the imaginary part of
     the cross-odd $P_{3g}$
     practically does not influences the form of the scattering amplitude. The imaginary part of
     the cross-even
     $P_{3g}$ exceeds the imaginary part of $P_{2g}$ when $-t>0.3$ GeV$^2$ and further determines
     the full Born term.

         The energy dependence of the  imaginary and real parts of the
          full Born term of the scattering amplitude is shown
         in Fig. 13. Note that the imaginary part grows with energy at small
         and decreases at large momentum transfer.
         The real part changes the sign and then grows at small
         momentum transfer but has a small energy dependence at large $t$.

         The energy dependence of the slope of the full Born term of the scattering amplitude
         is represented in Fig. 14.
         As $\hat{s}$ is complex, the slope has the real and imaginary parts too.
         The slope changes with $s$ most quickly  at small momentum transfer. The minimum
         change occurs in the region $-t \approx 0.8$ GeV$^2$.
         In Fig. 15, the energy dependence
         of the "basic" slope $B(s,t)$   [Eq.(\ref{B0})] is shown.
         Its real part has a simple form  at low energies with a growing  maximum
         at high energies that moves towards low $t$.
         However, the imaginary part has a more complicated energy dependence.
         The maximum of its  energy dependence occurs at a momentum transfer
         larger than that for the real part
         in the region  $0.4<-t<0.8$ GeV$^2$.
         The distinction of the form of the slope from the constant (linear) slope
         cannot be explained by the absence of the second Reggeons in the model.
         The second Reggeons have  real and imaginary parts of the same order
         with a large slope near $\alpha_{1} \approx 0.9$ GeV$^{-2}$.
         They  essentially change the $t$ dependence of the differential cross sections at low
         momentum transfer and especially in the CNI region.

         In order to use the unitarization procedure, it is necessary to transform
         the Born term of the scattering  amplitude in the impact-parameter presentation.
         As our Born term has the form factor in a complicated form
         compared to the simple exponential, this transform can be performed
          only by numerical integration,  [Eq.(\ref{chi})].
         Then, the standard eikonal representation is used to obtain the overlapping function
         in the impact-parameter representation, [Eqs. (\ref{FAmpl} and \ref{Gamma})].

\begin{figure}
  \includegraphics[width=.4\textwidth]{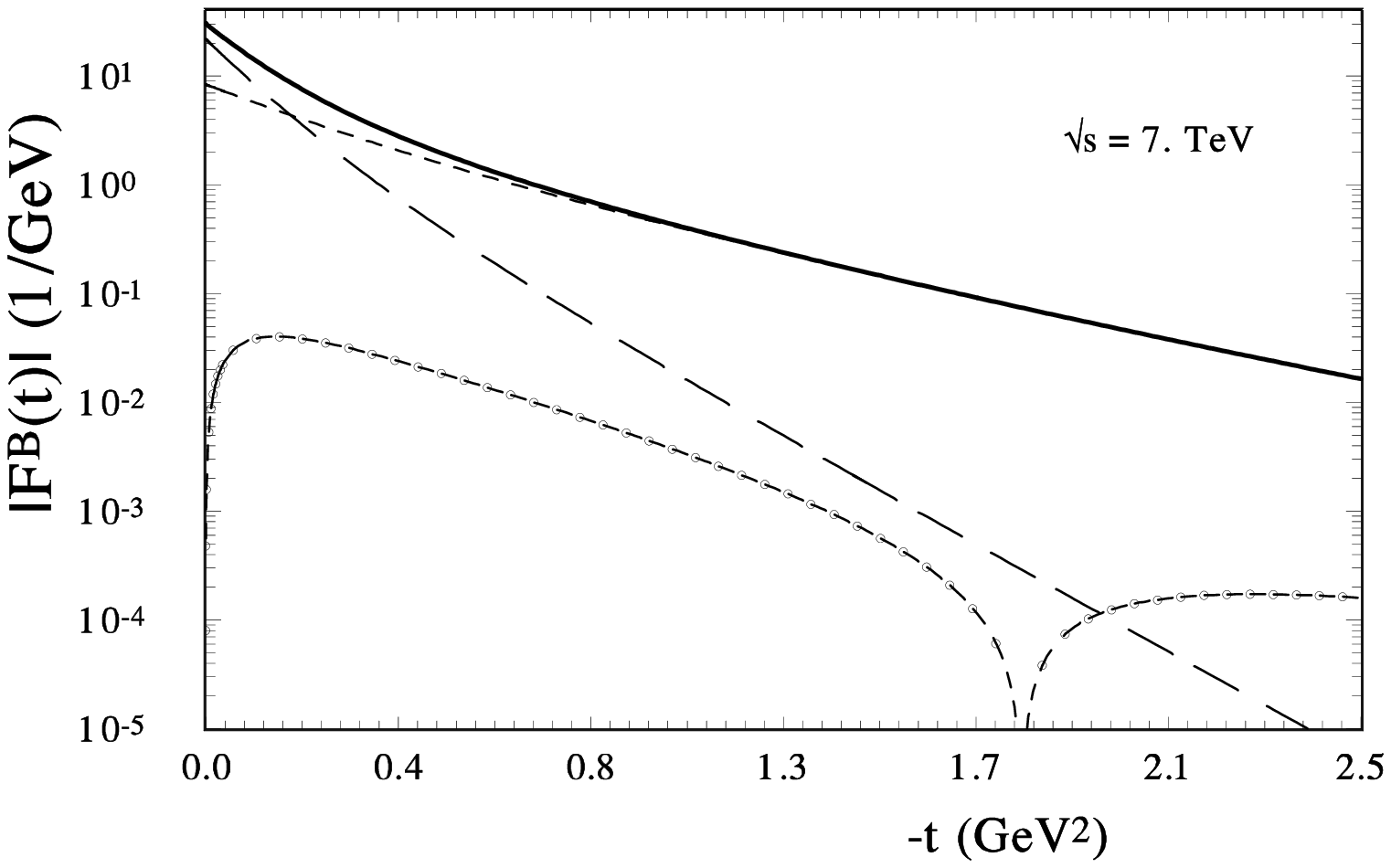}
      \includegraphics[width=.4\textwidth]{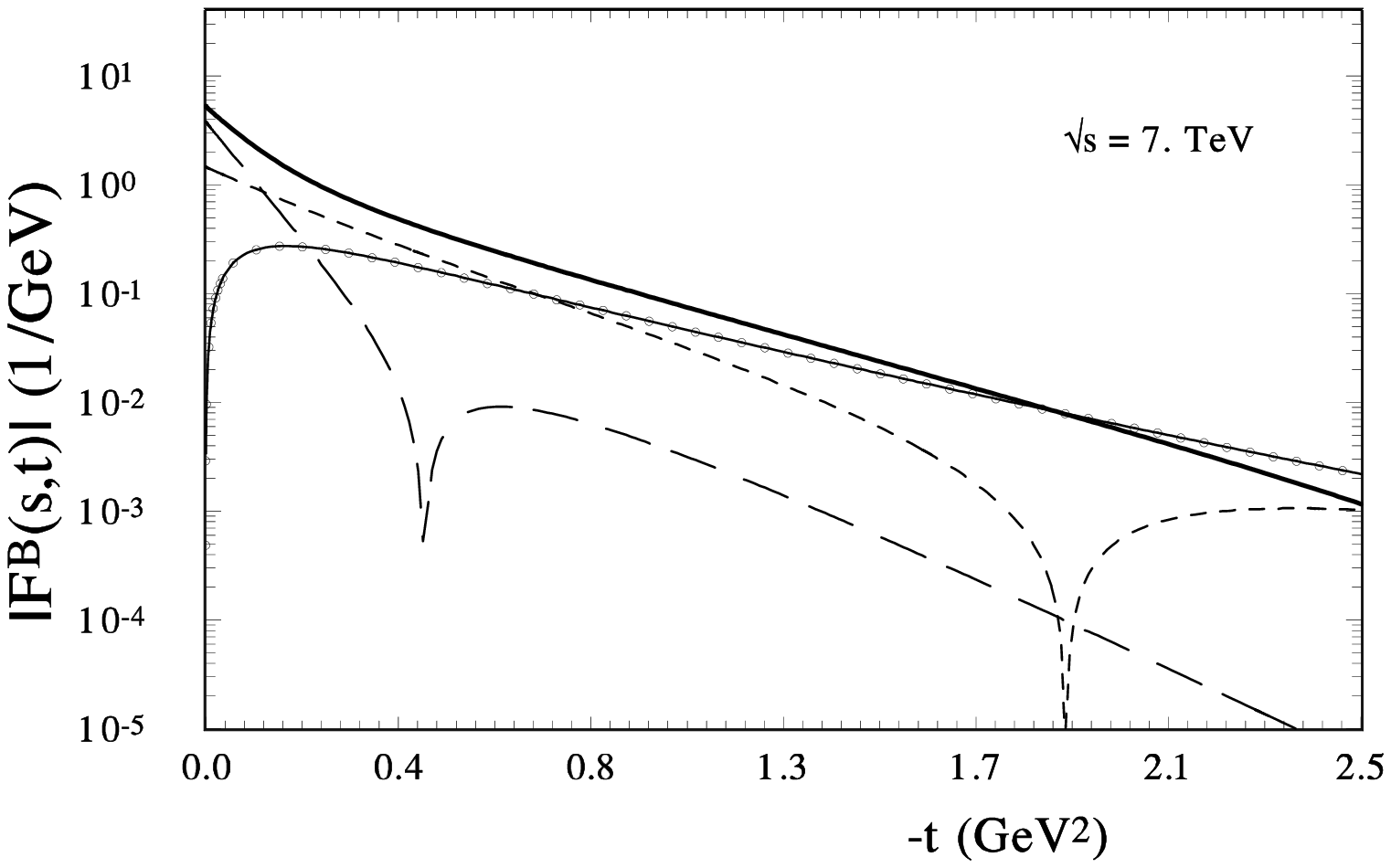}
  \caption{The magnitude of the Born parts of the $pp$ elastic scattering amplitudes
   at $\sqrt{s}=7$ TeV.
   (Top) The imaginary parts:  the sum of all parts (hard line),
    the contribution of  $P_{2g}$ (long-dashed line),
    the contribution of the cross-even $P_{3g}$  (dashed line),
   and the contribution of the cross-odd part of $P_{3g}$ (line with points).
  (Bottom) The same for the real parts of the Born amplitude.
   }
 \label{Fig12}
\end{figure}

         The energy dependence of the corresponding term of $\Gamma(s,b)$
         eq.(\ref{Gamma}) is represented in Fig. 16.
         One can see that the black disc limit
          is not reached at $\sqrt{s}=7$ TeV [where $Im  \ \Gamma(s,b=0)=0.93$]
          and even at $\sqrt{s}=14$ TeV [where $Im \ \Gamma(s,b=0)=0.95$].
          The real part of $\Gamma(b,s)$ (bottom panel of Fig. 16) is small and has some important influence
          only at large impact parameters.
         Hence, the behavior of the scattering amplitude will change to the saturation regime
         when  the energy  grows essentially above the LHC energies.
         It is necessary to take this into account when
         the cross sections are extrapolated from the accelerator energies
         to the energies reached in cosmic-ray experiments.

     The corresponding representations for the total, elastic and inelastic
     cross sections are
      \begin{eqnarray}
 \sigma_{\rm tot}(s) =
    \ 2 \int \ b  \Gamma_{\rm tot.}(s,b)  \ d b,
 \label{Gtot}
 \end{eqnarray}
   with $\Gamma_{\rm tot}(s,b) = Re\{ 1- \exp[ \chi(s,b)]\} $,
      \begin{eqnarray}
 \sigma_{\rm el}(s) =
    \ \int \ b   \Gamma_{\rm el}(s,b)  \ d b,
 \label{Gel}
 \end{eqnarray}
   with $\Gamma_{\rm el}(s,b)= Re\{ 1- \exp[\chi(s,b)]\}^2  $, and
      \begin{eqnarray}
 \sigma_{\rm inel}(s) =   \ \int \ b    \Gamma_{\rm inel}(s,b)  \ d b,
 \label{Ginel}
 \end{eqnarray}
   with $\Gamma_{\rm inel}(s,b) = Re \{ 1- \exp[ 2 \chi(s,b)]\}$.

\begin{figure}
  \includegraphics[width=.35\textwidth]{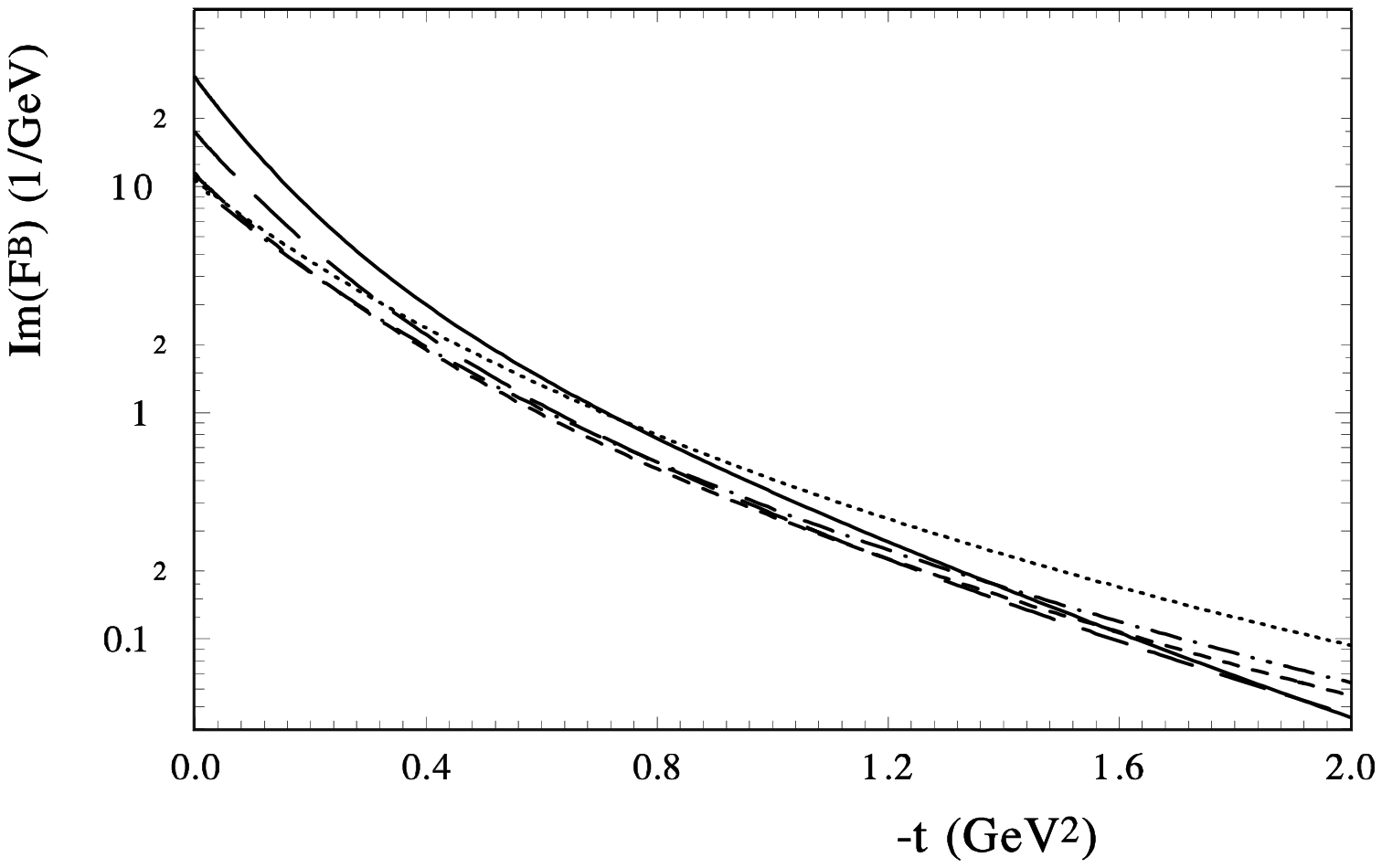}
      \includegraphics[width=.35\textwidth]{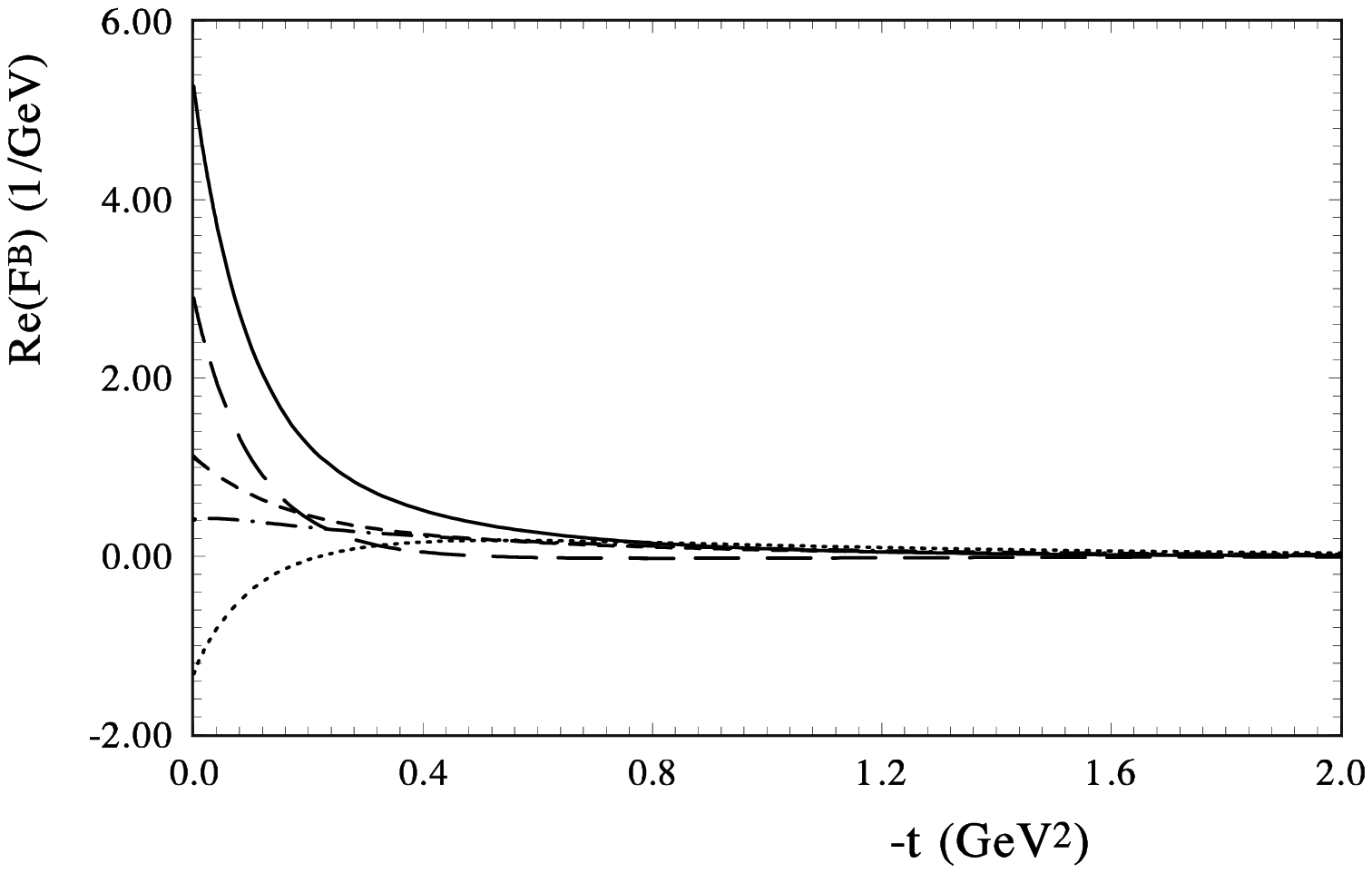}
  \caption{ (Top) panel]
   The imaginary part of the full Born amplitude of the the $pp$ elastic scattering amplitudes at
   $\sqrt{s}=7$ TeV (solid line),  $\sqrt{s}=541$ GeV (long-dashed line),
     $\sqrt{s}=52.8$ GeV (dashed line),
     $\sqrt{s}=27.4$ GeV (dashed-dotted line),  $\sqrt{s}=9.8$ GeV (dotted line).
 (Botton) The same for the real parts of the Born amplitude.
   }
  \label{Fig13}
\end{figure}

\begin{figure}
  \includegraphics[width=.35\textwidth]{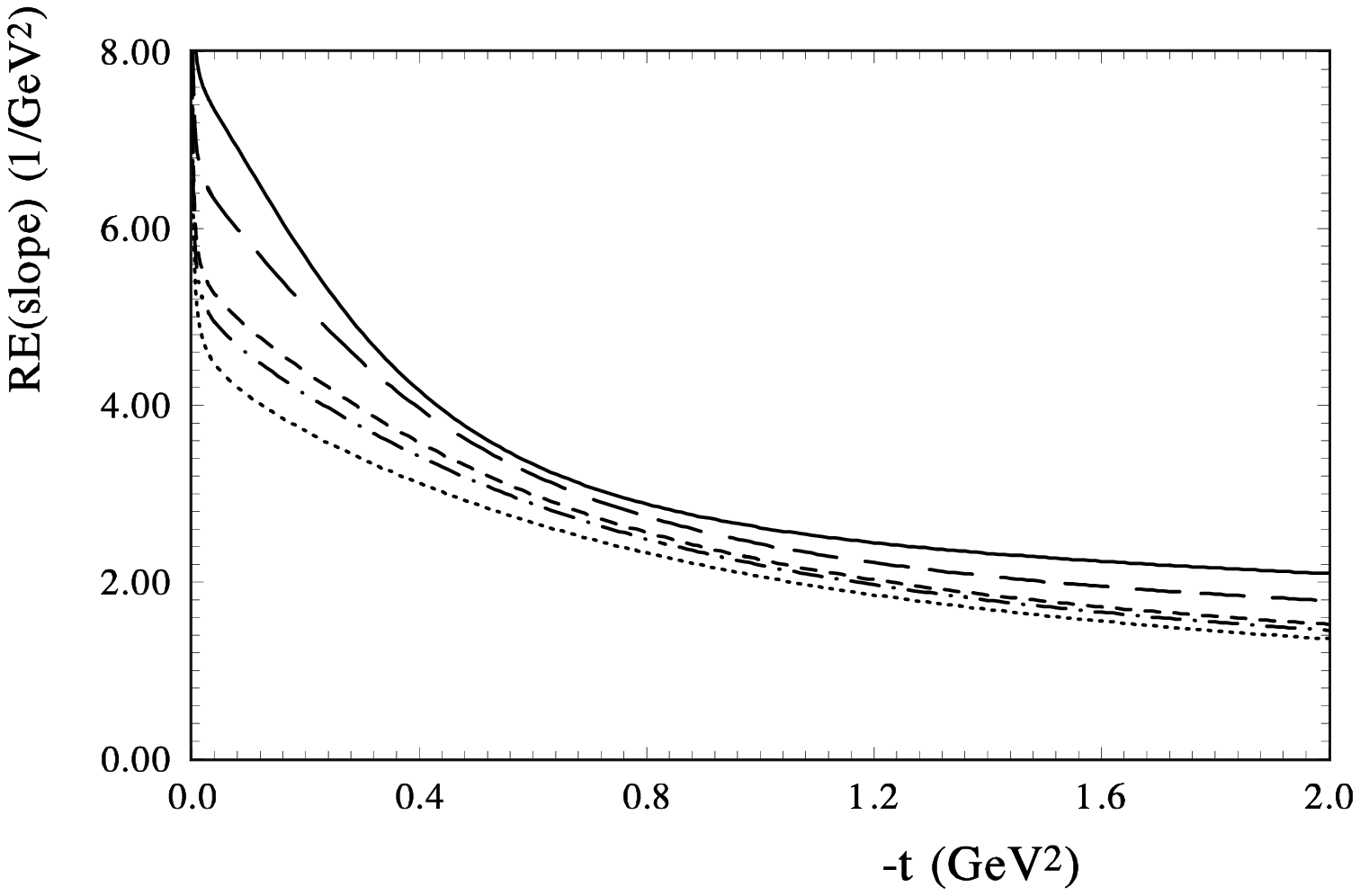}
      \includegraphics[width=.35\textwidth]{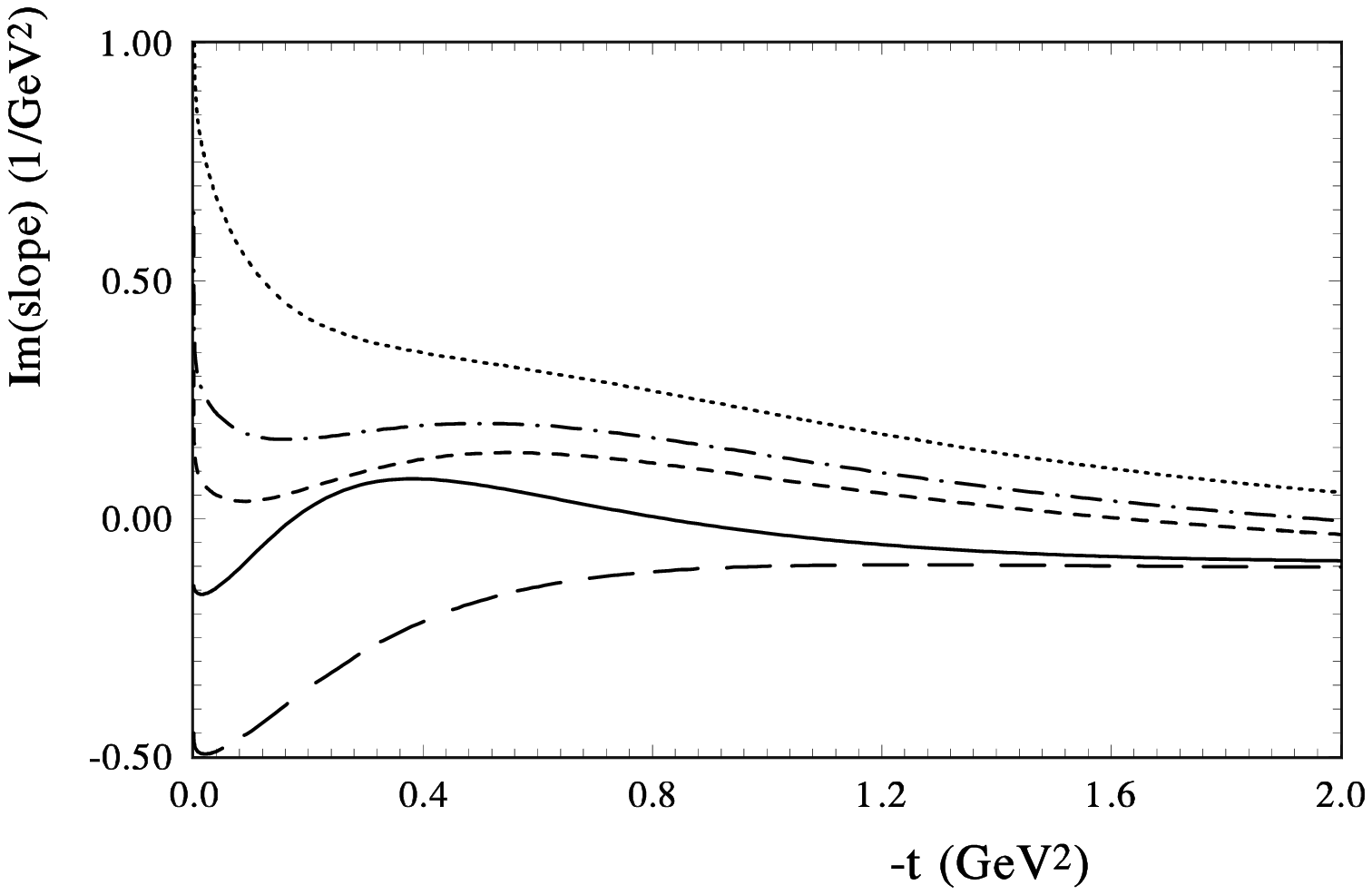}
  \caption{ (Top) The real part of the slope of the full Born amplitude of the the $pp$ elastic scattering amplitudes at
   $\sqrt{s}=7$ TeV (solid line),  $\sqrt{s}=541$ GeV (long-dashed line),
    $\sqrt{s}=52.8$ GeV (dashed line),
     $\sqrt{s}=27.4$ GeV (dashed-dotted line),  $\sqrt{s}=9.8$ GeV (dotted line).
 (Bottom) The same for the real parts of the Born amplitude.
   }
   \label{Fig14}
\end{figure}

\begin{figure}
  \includegraphics[width=.35\textwidth]{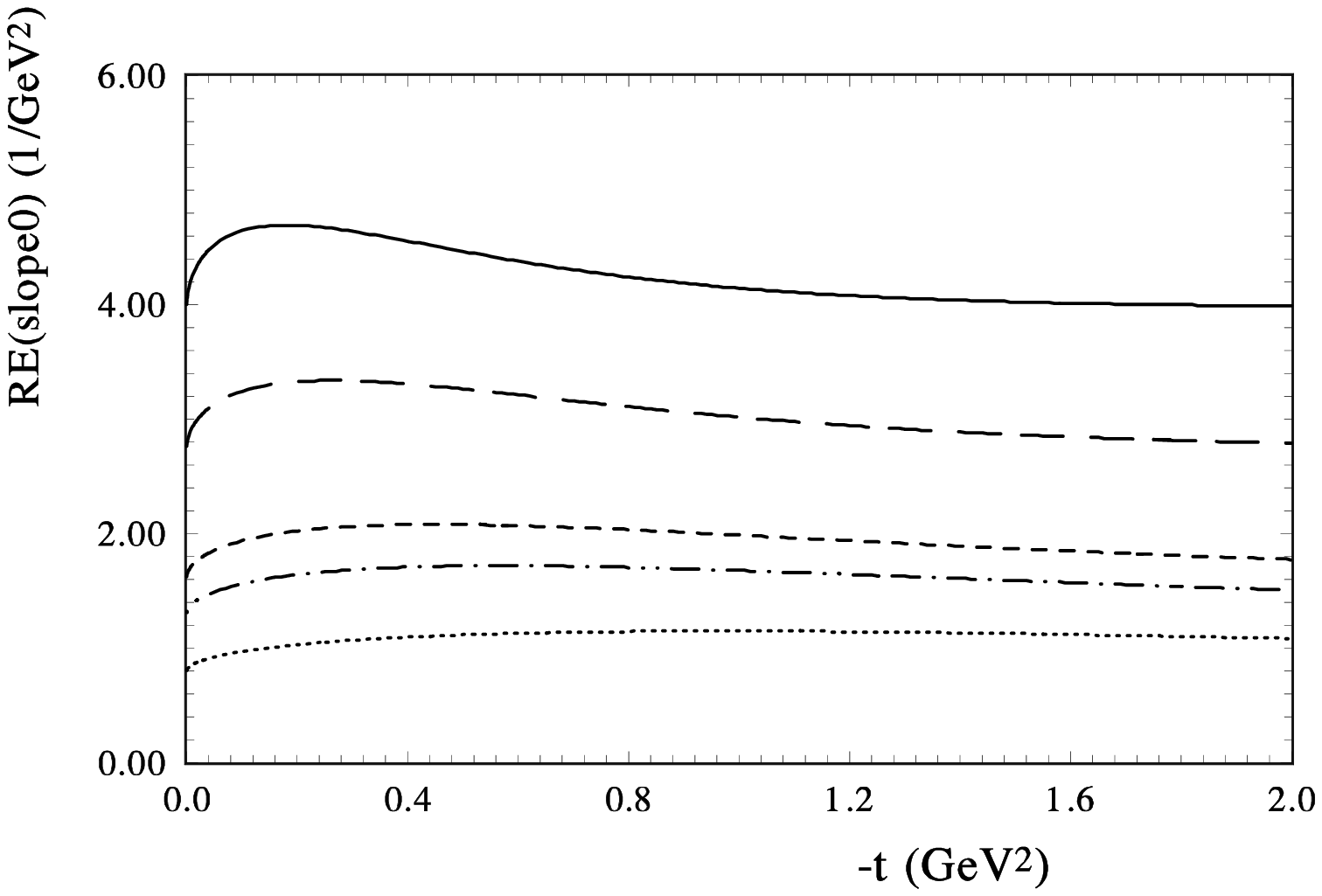}
      \includegraphics[width=.35\textwidth]{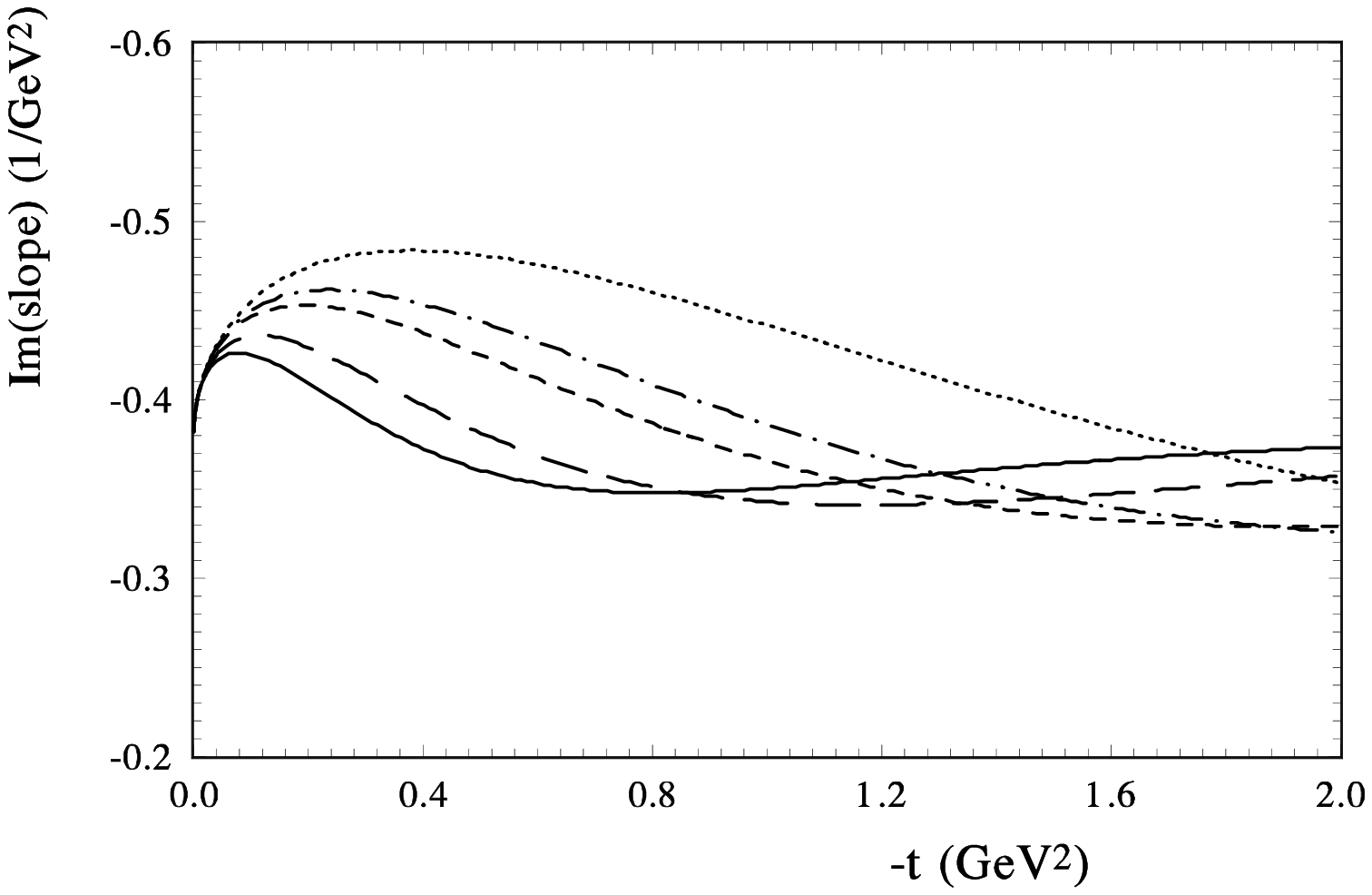}
  \caption{a) (Top)
   The real part of $B(s,t)$ [Eq.(\ref{B0})] of the the $pp$ elastic scattering amplitudes at
   $\sqrt{s}=7$ TeV (solid line),  $\sqrt{s}=541$ GeV (long-dashed line),
    $\sqrt{s}=52.8$ GeV (dashed line),
     $\sqrt{s}=27.4$ GeV (dashed-dotted line),  $\sqrt{s}=9.8$ GeV (dotted line),
 b) (Botton) The same for the imaginary parts of $B(,t)$ [Eq.(\ref{B0})].
   }
\label{Fig15}
\end{figure}

\begin{figure}
  \includegraphics[width=.45\textwidth]{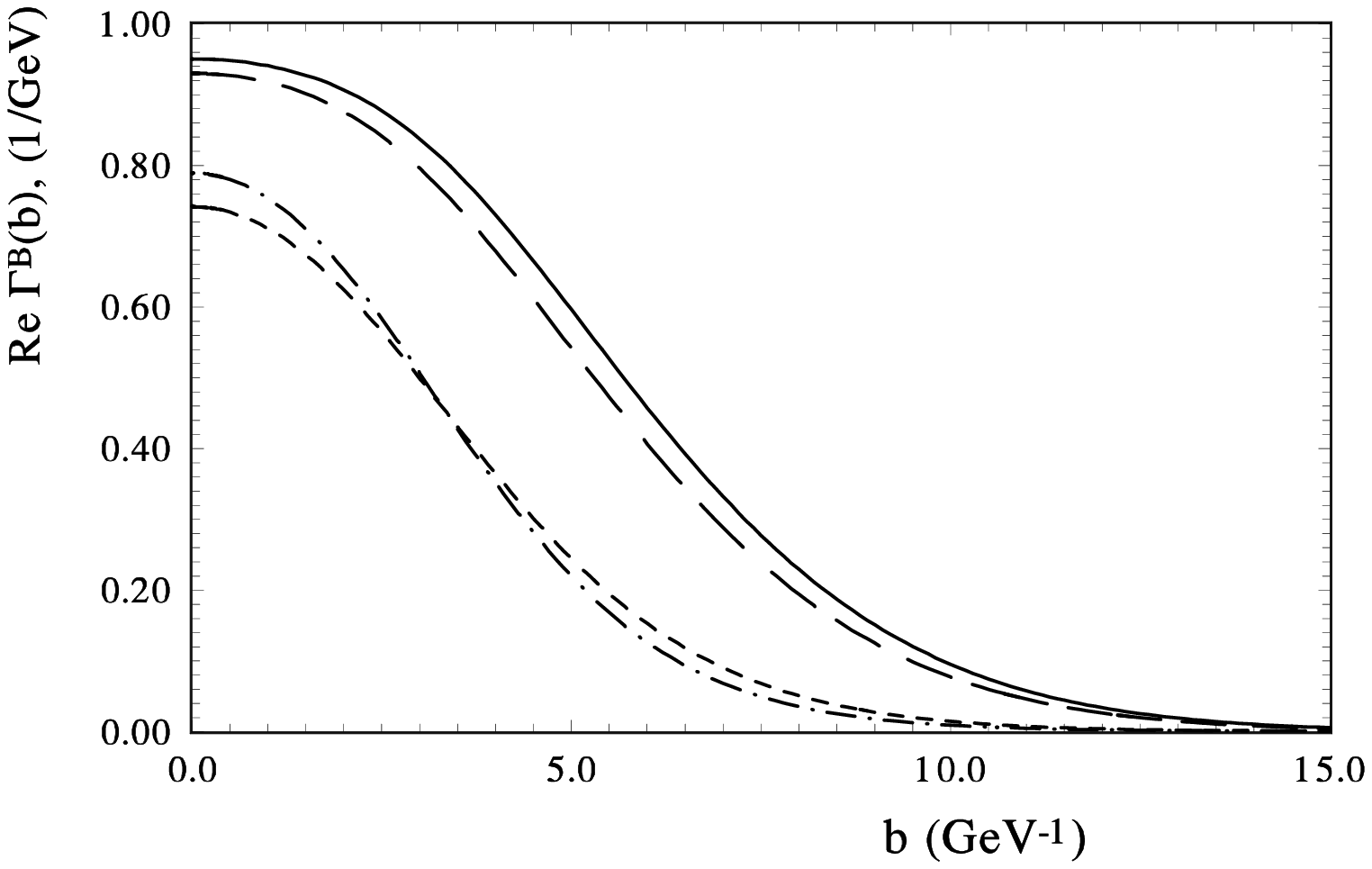}
    \includegraphics[width=.45\textwidth]{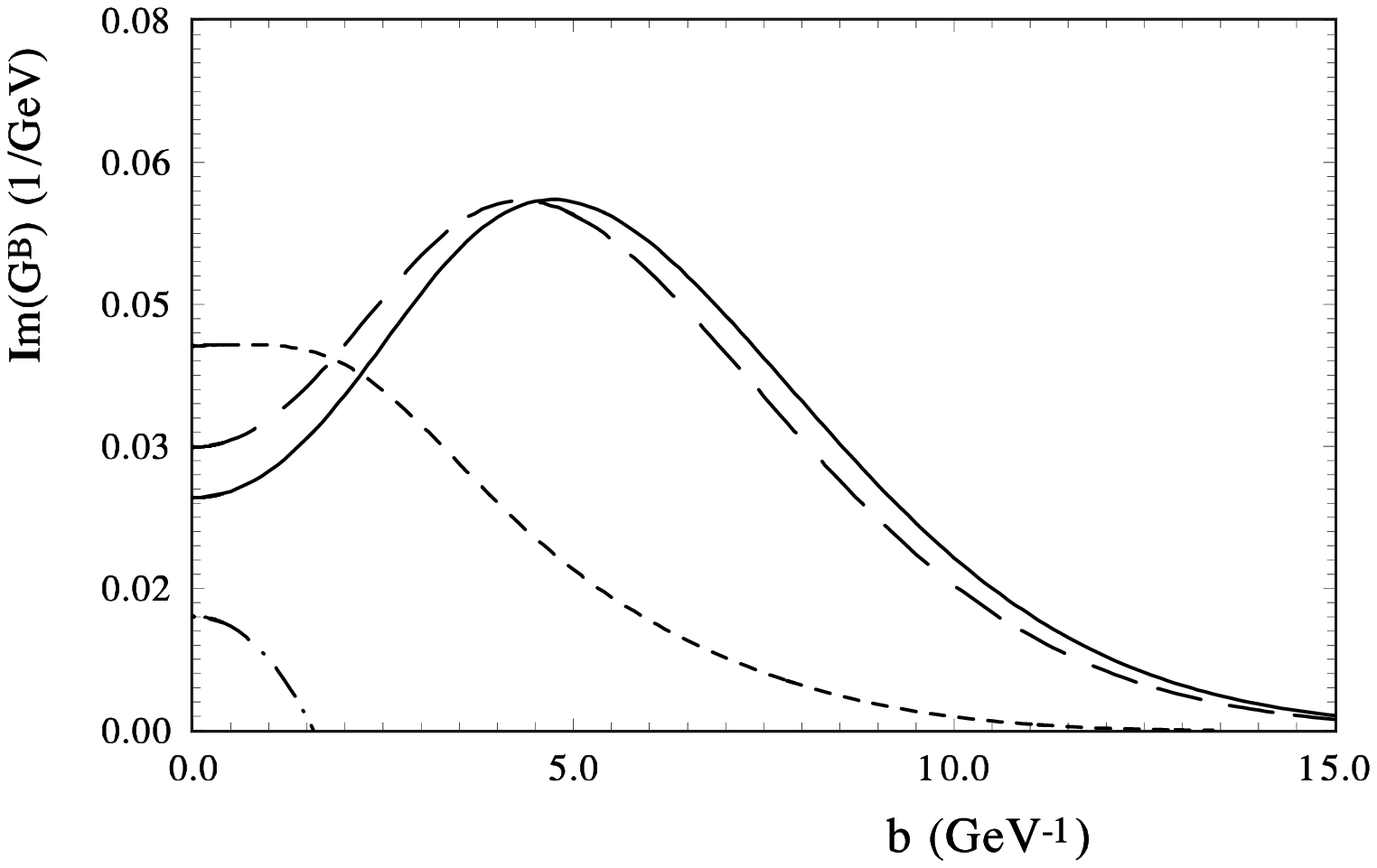}
  \caption{The overlapping function $\Gamma(s,b)$
    [for the real (top) and imaginary (bottom) parts]
    at $\sqrt{s}= 9.8$ GeV (dashed line), $\sqrt{s}= 52.8$ GeV (dash-dotted line),
    $\sqrt{s}= 7$ TeV (long dashed line),  $\sqrt{s}= 14$ TeV  (hard line).
   }
\label{Fig16}
\end{figure}

    The energy and impact parameter dependence of these values are represented in Fig. 17.
    The $\Gamma_{\rm inel}(s,b)$  saturates the unitarity bound up to $b=0.6$ fm at $\sqrt{s}=14$ TeV.
    However,  $\Gamma_{\rm el}(s,b)$ and the corresponding
    $\Gamma_{\rm tot}=\Gamma_{\rm el}+ \Gamma_{\rm inel}$
    do not reach the unitarity bound at this energy. In the middle panel of Fig. 17, we can see that such saturation
    takes place only at very large energy, $\sqrt{s}=100$ TeV. As a result, the maximal growth
    of
    $\Delta \Gamma_{\rm inel}(s_{2},s_{1})= \Gamma_{\rm inel}(s_{2}) - \Gamma_{\rm inel}(s_{2})$
    occurs at large impact parameters. The difference between
    $\Delta \Gamma_{\rm inel}(s_{2},s_{1})$,
    $\Delta \Gamma_{\rm el}(s_{2},s_{1})$ and $\Delta \Gamma_{\rm tot}(s_{2},s_{1})$
    of these values between $\sqrt{s}=14$ TeV and $\sqrt{s}=7$ TeV is shown in
    the bottom panel of Fig. 17.
    Though the size of the growth for the elastic and inelastic values is very similar,
    the maximum of the growth takes place at different impact parameters. The growth of inelastic processes has
     a peripheral character. However, it is due to the saturation of such processes in the central region.


\begin{figure}
  \includegraphics[width=.45\textwidth]{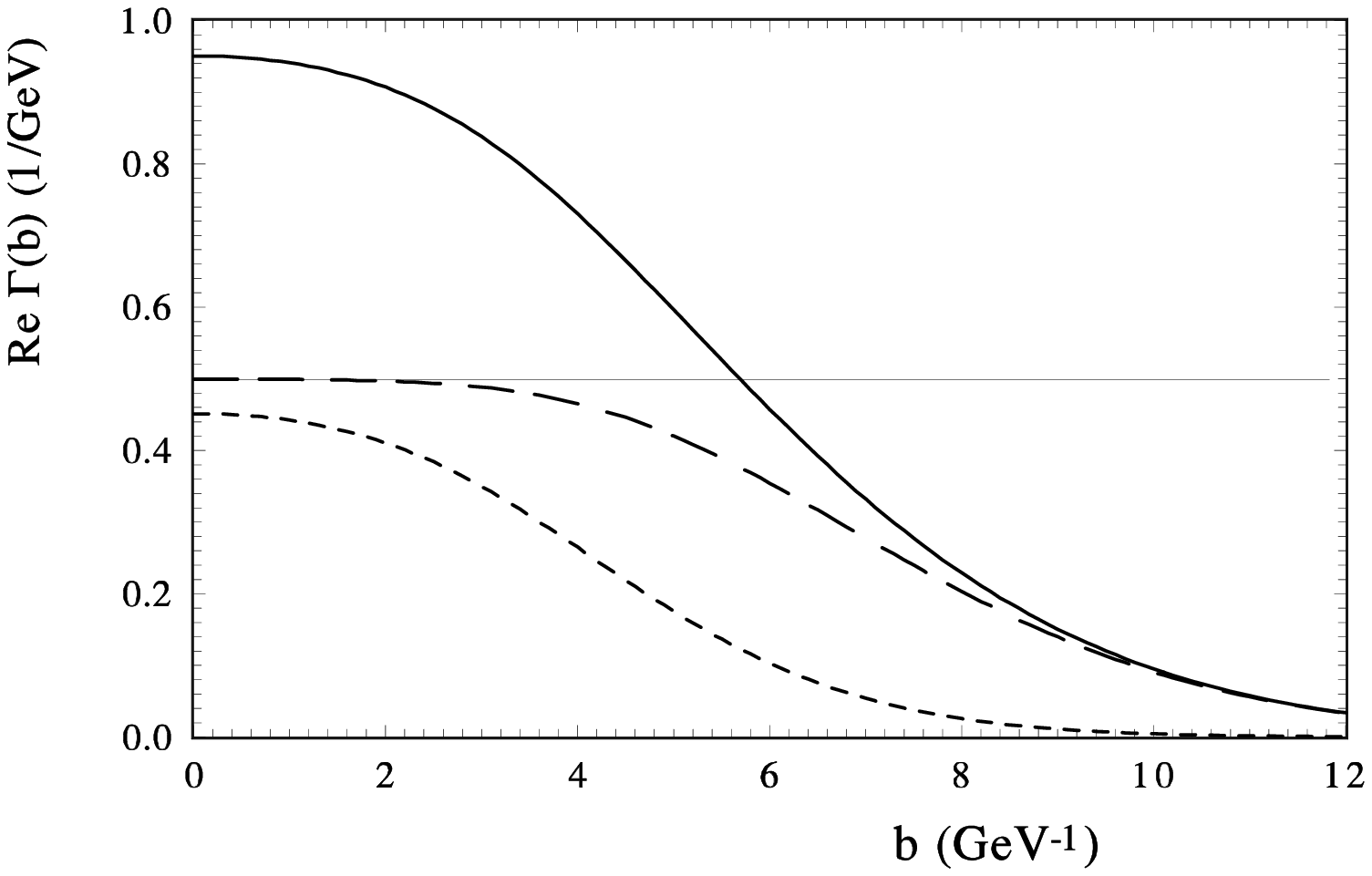}
   \includegraphics[width=.45\textwidth]{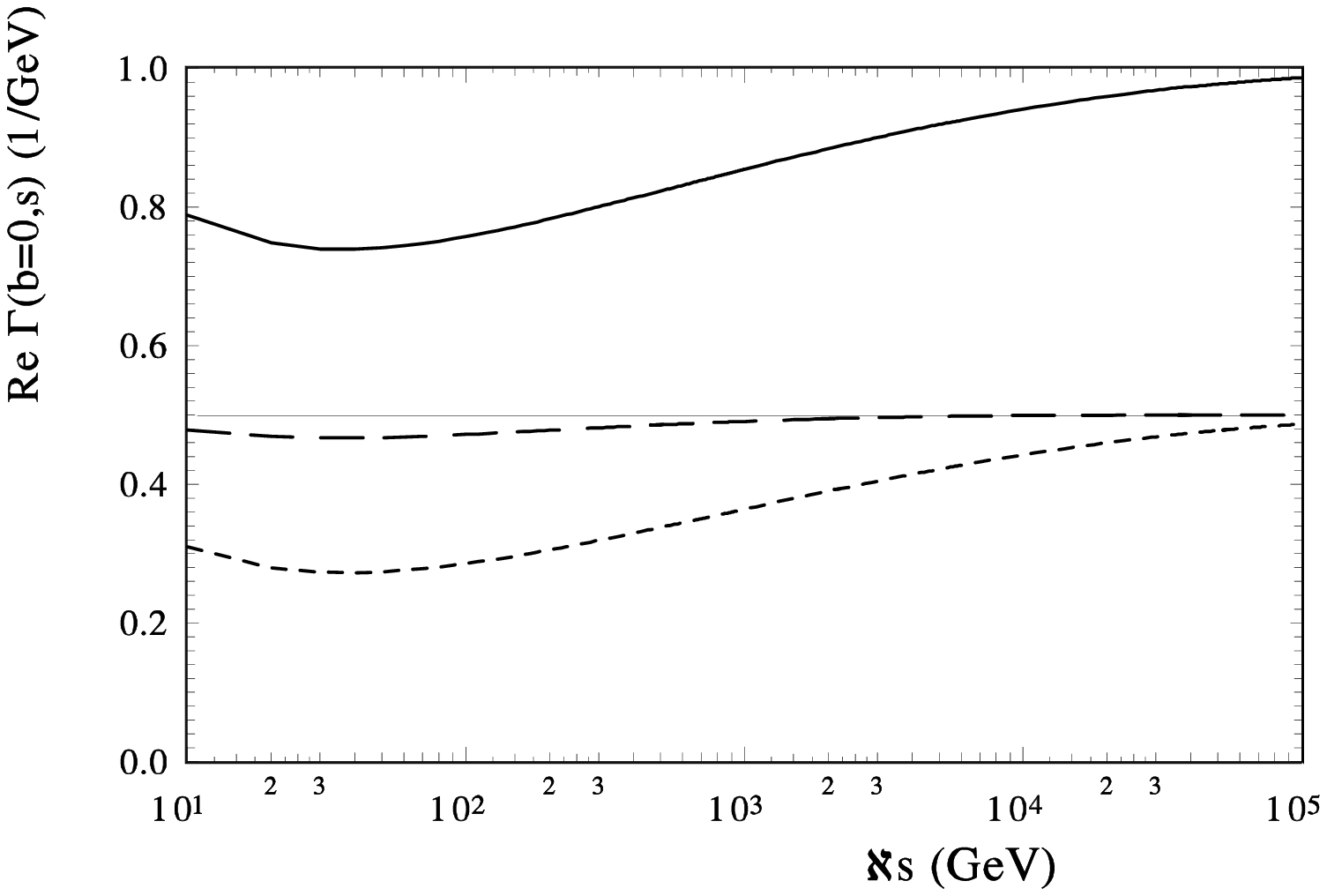}
   \includegraphics[width=.45\textwidth]{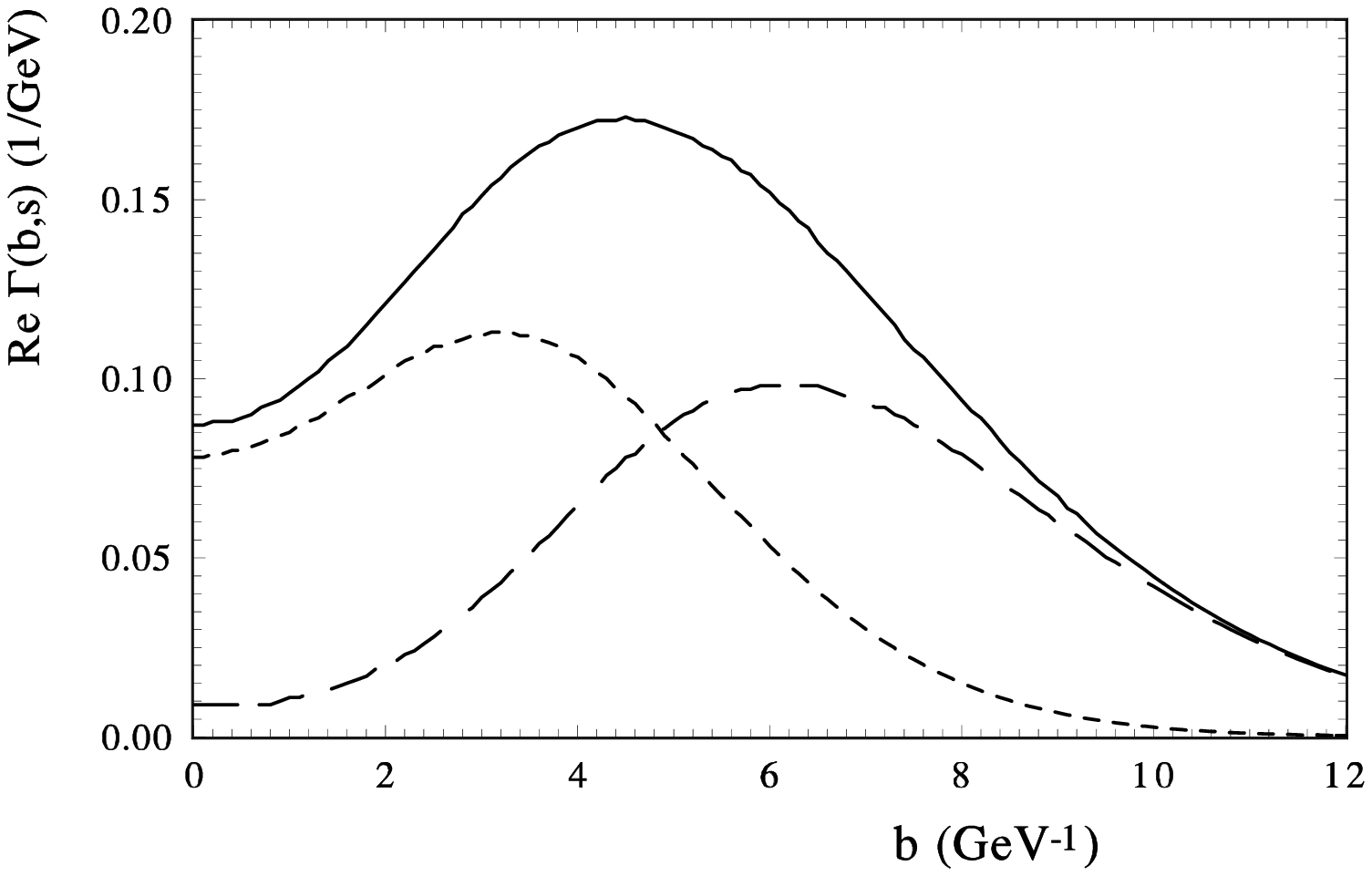}
  \caption{ (Top) $\Gamma(s,b)_{\rm tot}$(hard line), $\Gamma(s,b)_{\rm el}$(dashed line),
    $\Gamma(s,b)_{\rm inel}$(long dashed line) at $\sqrt{s}= 14$ TeV.
  (Middle) The energy dependence of $\Gamma(s,b=0)_{\rm tot}$(hard line), $\Gamma(s,b=0)_{\rm el}$(dashed line),
    $\Gamma(s,b=0)_{\rm inel}$(long dashed line).
     (Bottom) The differences between  $\Gamma(s,b)_{\rm tot}$(hard line), $\Gamma(s,b)_{\rm el}$(dashed line),
    $\Gamma(s,b)_{\rm inel}$(long dashed line) between $\sqrt{s}= 14$ TeV - $\sqrt{s}= 7$ TeV.
   }
  \label{Fig17}
\end{figure}

\begin{figure}
  \includegraphics[width=.45\textwidth]{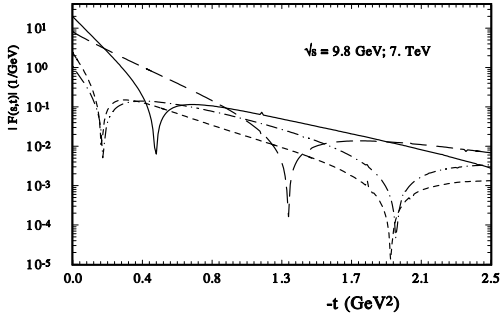}
  \caption{The magnitude of the $pp$ elastic scattering amplitudes after eikonalization
   of   the imaginary part and real parts for  for $\sqrt{s}= 9.8$ GeV (long-dashed line
   and dash-dotted line, respectively); and for $\sqrt{s}= 7$ TeV
  (hard line and dashed line,  respectively).
   }
   \label{Fig18}
\end{figure}

   The full elastic scattering amplitude is calculated
            by numerical integration [Eq.(\ref{FAmpl})].
        The magnitude of the real and imaginary parts of the full elastic scattering amplitude are represented in Fig. 18 for
          $\sqrt{s}=9.8$ GeV to  $\sqrt{s}=7$ TeV.
          The imaginary part changes by moving
          its zero from $-t  = 1.35$ GeV$^2$ to  $-t  = 0.5$ GeV$^2$.
           The magnitude of the real part
          has  zeroes in the examined region of $t$ and  mostly  changes its size,
          but conserves its form.

\begin{figure}
  \includegraphics[width=.45\textwidth]{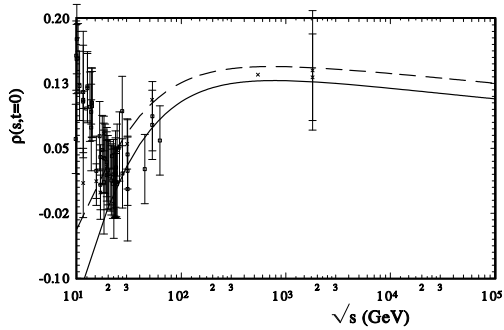}
  \caption{The energy dependence of $\rho(s,t=0)$ (the ratio of the real and imaginary parts of the scattering   amplitude for  $pp$ (solid line) and   $\bar{p}p$ (dashed line) scattering.
      We also show the experimental data for $pp$ (squares) and  $\bar{p}p$ (crosses) scattering
   from Ref. \cite{data-Sp}.
   }
   \label{Fig19}
\end{figure}

\begin{figure}
  \includegraphics[width=.45\textwidth]{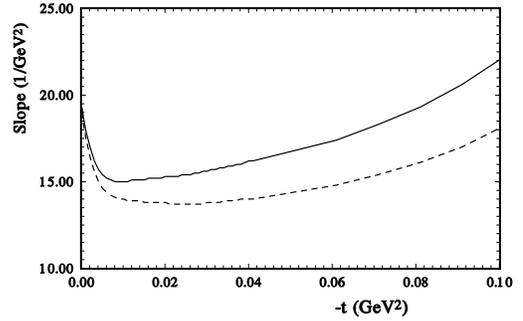}
   \includegraphics[width=.45\textwidth]{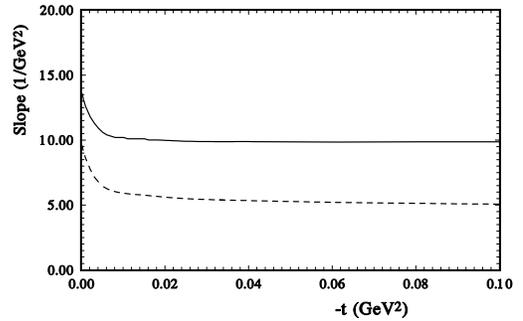}
    \includegraphics[width=.45\textwidth]{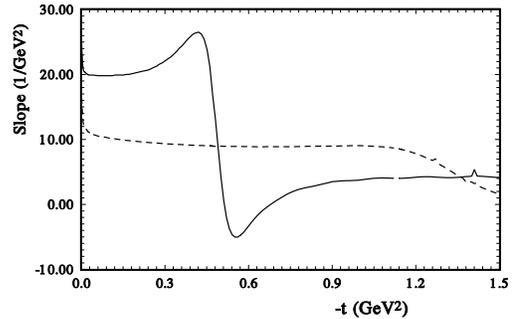}
  \caption{The slope of the full scattering amplitude
    at $\sqrt{s}= 9.8$ GeV (dashed line) and
    $\sqrt{s}= 7$ TeV (hard line), for the real parts (top),
     the imaginary parts (middle), and the slope of the differential
    cross sections (bottom).
   }
   \label{Fig20}
\end{figure}

          It should be noted that the real part is negative at small
          momentum transfer at $\sqrt{s}=9.8$ GeV.
   The energy dependence of $\rho(s,t=0)$ (the ratio of the real part to the imaginary part of the scattering amplitude)
   is shown in Fig. 19. Note that we do not include the experimental data on $\rho(s,t=0)$
   and $\sigma_{tot}(s)$  in the fitting procedure.

      In Fig. 20, the slopes of the hadronic part of the full elastic scattering amplitude
       are represented at $\sqrt{s}=9.8$ GeV and  $\sqrt{s}=7$ TeV.
      Obviously, the difference between the slopes at small momentum transfer  is only in the size,
      though we examined such different energies.
       The nonlinear behavior of the slope of the Born term of the scattering amplitude is
       very weakly reflected in the form of the slope of the eikonalized amplitude,
        but it essentially influences  the fitting procedure.
       In a larger region of the momentum transfer
      the slope of the differential cross section has a significant difference
      for the low and high energies (bottom panel of Fig. 20). This is the result of the eikonalization procedure
      which is reflected in the position of the diffraction minimum.

  The comparison of the energy dependence of the model calculations
  of the slope of the differential cross sections is represented in Fig. 21.
  In Ref. \cite{Bloch}, the slope was determined as

      \begin{eqnarray}
 B_{el}(s) =
    \frac{ \int \ d^{2}b \ b^2 \ \Gamma(s,b)}{  \int \ d^{2}b \ b \ \Gamma(s,b)}  ;
 \label{BelBlock}
 \end{eqnarray}

    This gives the slope at $t=0$, but the experimental data for the slope are
   obtained at small momentum transfer and beyond the Coulomb-hadron interference region
   and in some region of $t$. So, we used the standard determination of the slope,
      \begin{eqnarray}
 B_{el}(s) =
   Log[ \frac{ d\sigma/dt|_{t_{1}} }{ d\sigma/dt|_{t_{2}} } ]/[|t_{2}|-|t_{1}|   ] ;
 \label{Bel}
 \end{eqnarray}

\begin{figure}
   \includegraphics[width=.45\textwidth]{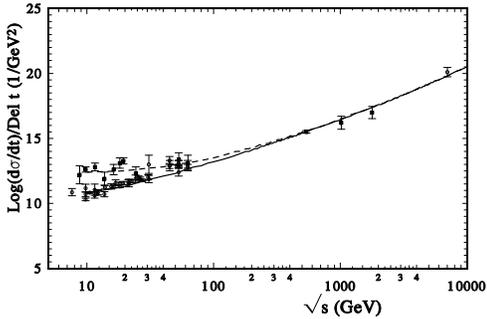}
  \caption{The energy dependence of the forward elastic  slope [Eq. (\ref{Bel})]  compared to the existing
  experimental data for  $pp$  (hard line and open circles) and  $p\bar{p}$ (dashed line and squares)
   scattering.
   }
   \label{Fig21}
\end{figure}

       In the different experimental data, $t_{1}$ and  $t_{2}$ are different. We take
      some middle points:  $-t_{1}=0.04$ GeV$^2$ and  $-t_{2}=0.05$ GeV$^2$.
      The experimental data have large errors. However,
      the energy dependence of our calculations for the most part coincides with
      the energy dependence of the experimental data.

\section{Conclusions}

  We have presented a new model of the hadron-hadron interaction at high energies.
  The model is very simple with regards to the number of parameters and functions.
  There are no artificial functions or cuts that bound the separate
  parts of the amplitude by some region of momentum transfer or energy.
  One of the most remarkable properties is that the real part of the
  hadron scattering amplitude is determined only by complex energy $\hat{s}$
  that satisfies the crossing symmetries.

 The new HEGS
  model  gives a quantitative     description of the elastic nucleon scattering at high energy  with only five fitting high-energy parameters.
 Our model of the GPDs leads to a good description of the proton and neutron  electromagnetic form factors and their elastic scattering simultaneously.
  A successful description  of the existing experimental data by the model shows that
   the elastic scattering  is determined by the generalized structure of the hadron.
The model leads to a  coincidence of the model calculations with the preliminary data at 8 TeV.
   We found that the  standard eikonal approximation \cite{Unit-PRD}   works perfectly well from
  $\sqrt{s}=9$  GeV up to  8 TeV.
The extended variant of the model shows the contribution of the "maximal" Odderon with specific
   kinematic properties
  and does not show a visible contribution of the hard Pomeron, as in Ref. \cite{NP-HP}.

\begin{table}
 \caption{The obtained \cite{PDG} and predicted sizes of the $\sigma_{\rm tot}(s)$, mb and $\rho(t=0,s)$ }
\label{Table-5}
\begin{center}
\begin{tabular}{|c|c|c|c|} \hline
 $\sqrt{s}$, GeV &$\sigma_{\rm tot-exp}$   & $\sigma_{\rm tot}$ & $\rho(t=0,s)$  \\  \hline
 $19.42$ & $38.98\pm0.4$    & $39.58\pm0.8$  &  $-0.005\pm0.0006$  \\
 $22.96$ & $39.42\pm0.4$    & $39.89\pm0.8$  &  $-0.005\pm0.0006$  \\
 $52.8$ & $42.85\pm0.7$    & $43.15\pm0.5$  & $0.074\pm0.005$      \\
 $541$  & $62.72\pm0.2$    & $62.72\pm0.2$  & $0.128\pm0.005$   \\
$1800$  & $77.3\pm0.38$    & $77.3\pm0.38$  &$ 0.127\pm0.02$  \\
$7000$  & $98.0\pm2.6$    & $97.16\pm0.5$  & $ 0.121$  \\
$7000$  & $96.4\pm 2.$    & $97.16\pm0.5$  & $ 0.121$  \\
$8000$  & $101.\pm2.1 $    & $99.4\pm0.5$  &$ 0.12$  \\
$14000$ & $104\pm26. $   & $108.76\pm0.5$ &$ 0.1176$  \\
$30000$ & $120.\pm 15$    & $122.7\pm0.5$  &$ 0.11$  \\
$57000$ & $133 \pm 23 $    & $135.4\pm0.5$  &$ 0.11$  \\
 \hline
\end{tabular}
\end{center}
\end{table}

\begin{table}[b]
 \caption{The sizes of the cross sections and the ratio of the $\sigma_{\rm el}/\sigma_{\rm inel}$ at $7$ TeV }
\begin{center}
\begin{tabular}{||c|c|c|c|c|} \hline
    & $\sigma_{\rm tot}$ & $\sigma_{\rm el}$ &  $\sigma_{\rm inel}$  & $R_{\rm el/inel}$  \\ \hline
 $\rm HEGS_{0}$ & $95.1\pm0.8$ &  $24.1\pm0.8$ &  $65\pm0.8$ & $0.37$ \\
 $\rm HEGS_{1}$  & $97.2\pm0.5$  & $24.3\pm0.5$  & $66\pm0.5$ & $0.37$     \\  \hline
 $\rm TOTEM$ & $98.3\pm2.9 $  & $24.8 \pm 1.3$  & $73.5 \pm 1.9$   & $0.34 $    \\
$\rm ATLAS$ & $95.35 \pm 1.36 $  &$ 24.0 \pm0.6$  & $71.34\pm0.9$  & $0.34 $ \\
 \hline
\end{tabular}
\end{center}
  \end{table}

\begin{table*}
 \caption{The properties of some models ( $BSW_{1}$ \cite{BSW}, $BSW_{2}$ \cite{Bourrely-14}, $AGN$ \cite{AGN},
           $MN$ \cite{MN}, $HEGS_{0}$ \cite{HEGS-JEP12},$HEGS_{1}$ -present  )}
\label{Table-7}
\begin{center}
\begin{tabular}{||c|c|c|c|c|c|c|} \hline
  & $\rm BSW_{1}$ & $\rm BSW_2$ &  AGN  & MN & $\rm HEGS_0$ & $\rm HEGS_1$  \\\hline
 $N_{\rm exp}$ & $369$ &  $955$ &  $1728+178$& $2654+238$   &$980$ & $3416$ \\
 $ n_{\rm par}$  & $7+\rm Regge$  & $11+\rm Regge$  & $35$ & $43$  &  $3+2$ &$5+4$   \\
 $\sqrt{s}$, GeV     & $24-630$ & $13.4-1800$ & $9.3-1800 $ & $5-1800$ & $52-1800$  &  $9-8000$  \\
 $\Delta t$, GeV$^2$ & $0.1-2.6$  & $0.1-5.  $ & $0.1-2.6$ &  $0.1-15$ & $8.7 \ 10^{-4}-10 $  & $3.7 \ 10^{-4}\div15$ \\
$(\sum \chi^2)/N $ & $4.45$ & $1.95$ & $2.46$   & $1.23$  &  $1.8$    &$1.28$  \\
 \hline
\end{tabular}
\end{center}
  \end{table*}

The  slope of the differential cross sections at small momentum transfer has a small
      peculiarity and has the same properties in the whole examined energy
      region. Such a uniform picture for the slope gives the possibility of further research
        into small peculiarity of  different cross sections, such as possible oscillations \cite{LRPot-Dif10}.
   Note that  we did not see the contributions of the second Reggeons with a large slope and intercept above 0.5 in the examined  energy region.

   The obtained value of the total cross sections  $\sigma_{\rm tot}(s)$ and parameter $\rho(s,t=0)$ are shown in Table V.
   At low energies the model calculation of $\sigma_{\rm tot}(s)$ corresponds to the experimental data.
     The inclusion in our fit of the data of the TOTEM Collaboration increases the
       $\sigma_{\rm tot}$  at $\sqrt{s}=7$ TeV  from  $95$ mb \cite{HEGS-JEP12}    to $97$ mb,
       which can be compared with recent data \cite{TOTEM-8nexp}.

       In Table VI,  our model calculations  at $\sqrt{s}=7$ TeV are compared
       with experimental data obtained at the LHC by  different experimental collaborations.
       On the whole, the model calculations correspond to the existing experimental data.
       In our opinion, the experimental result of the ATLAS Collaboration
        $\sigma_{\rm tot}(s)$ is preferable.

  In  Table VII, the comparison of our model with some others are shown.
  The first is the old Bourrely-Soffer-Wu  $(BSW_{1})$ model \cite{BSW}.
  It has a small  number of fitting parameters (seven), but it also takes
   into account the second Reggeon's contributions.
  We have  already noted that in this model the form factor is approximated by some function
  and it represents  the average between the electromagnetic and  matter form factors.
  The recent development of this model is represented as $\rm BSW_{2}$ \cite{Bourrely-14}.
  In this variant the number of  experimental points  increases,
  but the number of fitting parameters  increases as well.
  Both of  the models (ours and BSW)  used the standard eikonal unitarization procedure.
  Another example of the model (which does not use the unitarization procedure)
 is  based for the most part on Ref. \cite{GLN90}. Its recent variants are presented
  as AGN[\cite{AGN}]   and MN [\cite{MN}].
  The work is based on many different forms of the parts of the elastic scattering amplitude
  with many additional artificial functions. It examined  a large number of experimental data points
  but did not include the Coulomb-hadron interference region.
  The model includes in the fitting procedure
  the experimental data of the total cross sections $\sigma_{\rm tot}(s)$ and parameter $\rho(s,t=0)$.
  Such an inclusion decreases the total $\chi^2$.
  Both  models summed the statistical and systematic errors and have
   $\chi^2/N$ $2.46$ and $1.23$,  respectively.
  The obtained $\chi^2$ is minimal in the model MN
  but with a huge number of fitting parameters and the inclusion of  additional artificial functions.
  Only our model takes into account the whole region of  momentum transfer
  ($3.75 \ 10^{-4} \geq |t| \geq 15$ GeV$^{2}$, which includes the high-precision experimental data
  in the Coulomb-hadron interference region.


    A small number of  fitting parameters will make it possible to explore  some fine additional
    effects like possible oscillations of the scattering amplitude and
    find  some corrections to the standard eikonal unitarization procedure.

\vspace{0.5cm}

{\center{\bf Acknowledgments} } \\
 The authors would like to thank J.-R. Cudell and O.V. Teryaev
   for fruitful   discussions of some questions   considered in this paper.

\end{document}